\documentclass[a4paper,12pt]{article}

\usepackage{bbm}
\usepackage[utf8]{inputenc}
\usepackage{placeins}
\usepackage{amssymb,amsmath,amsfonts}
\usepackage[colorlinks,linkcolor=purple,citecolor=teal]{hyperref}
\usepackage{dsfont}
\usepackage{color}
\usepackage{graphicx}
\usepackage{hyphenat}
\usepackage{wrapfig}
\usepackage{empheq}
\usepackage{textcomp}
\usepackage[caption=false]{subfig}
\usepackage{rotating}
\usepackage{wrapfig}
\usepackage{pdfpages}
\usepackage{verbatim}
\usepackage{setspace}
\usepackage{array}
\usepackage{caption}
\usepackage[pdf]{pstricks}
\usepackage{upgreek}
\usepackage{cmll}
\usepackage{latexsym}
\usepackage{braket}
\usepackage{cite}
\usepackage{epsfig}
\usepackage[left=2.4cm,top=3.3cm,right=2.4cm,bottom=3.3cm,bindingoffset=0cm]{geometry}
\usepackage[titletoc,page]{appendix}
\usepackage{multicol}
\usepackage{youngtab}

\newcommand{\beq}{\begin{eqnarray}}
\newcommand{\eeq}{\end{eqnarray}}
\newcommand{\bea}{\begin{eqnarray}}
\newcommand{\eea}{\end{eqnarray}}
\newcommand{\be}{\begin{equation}}
\newcommand{\ee}{\end{equation}}

\newcommand{\calR}{\mathcal{R}}
\newcommand{\rmc}{\mathrm{c}}

\newcommand{\rmF}{\mathrm{F}}

\newcommand{\rmS}{\mathrm{S}}
\def\brc{\langle}
\def\ckt{\rangle}
\def\const{{\rm const}}
\def\de{\partial}

\def\nn{\nonumber}

\def\Tr{\qopname\relax o{Tr}}

\numberwithin{equation}{section}

\def\bxi{{\bar{\xi}} }
\def\su{$ \phantom{{{{{\yng(1)}}}}}\!\!\!\!\!\!\!\!$}
\def\sbuu{$\phantom{{{{\bar{\yng(1,1)}}}}}\!\!\!\!\!\!$}
\def\sbu{$\phantom{{\bar{{{\yng(1)}}}}}\!\!\!\!\!\!\!\!$}
\def\sbbuu{$\phantom{{{\bar{\bar{\yng(1,1)}}}}}\!\!\!\!\!\!\!\!$ }
\def\sbbu{ $\phantom{{{\bar{\bar{\yng(1)}}}}}\!\!\!\!\!\!\!\!$ }

\numberwithin{equation}{section}

\def\sp{-\!\!}

\begin{document}

\title{
\vskip 20pt
\bf{Strong anomaly  and  \\   phases of  chiral gauge theories}
}

\vskip 60pt  

\author{  
Stefano Bolognesi$^{(1,2)}$, 
 Kenichi Konishi$^{(1,2)}$, Andrea Luzio$^{(3,2)}$    \\[13pt]
{\em \footnotesize
$^{(1)}$Department of Physics ``E. Fermi", University of Pisa,}\\[-5pt]
{\em \footnotesize
Largo Pontecorvo, 3, Ed. C, 56127 Pisa, Italy}\\[2pt]
{\em \footnotesize
$^{(2)}$INFN, Sezione di Pisa,    
Largo Pontecorvo, 3, Ed. C, 56127 Pisa, Italy}\\[2pt]
{\em \footnotesize
$^{(3)}$Scuola Normale Superiore,   
Piazza dei Cavalieri, 7,  56127  Pisa, Italy}\\[2pt]
\\[1pt] 
{ \footnotesize  stefano.bolognesi@unipi.it, \ \  kenichi.konishi@unipi.it,  \ \  andrea.luzio@sns.it}  
}
\date{}

\vskip 6pt

\maketitle

\begin{abstract}

We present a simple argument which seems to favor, when applied to a large class of strongly-coupled chiral gauge theories,   
a dynamical-Higgs-phase scenario, characterized by certain bifermion condensates. 
 Flavor symmetric confining vacua described in the infrared by a set of baryonlike massless composite fermions  
 saturating the conventional 't Hooft anomaly matching equations, appear instead disfavored.  
 Our basic criterion is that it should be possible to write a strong-anomaly effective action, analogous to the one used in QCD 
 to describe the solution of the $U(1)_A$ problem in the low-energy effective action,    
  by using the low-energy 
degrees of freedom in the hypothesized infrared theory.   
We also comment on some well-known ideas such as the complementarity and the large $N$  planar dominance in the context of these chiral gauge theories. 
  Some striking analogies and contrasts between the massless QCD  and chiral gauge theories seem to emerge from this discussion.

\end{abstract}

\newpage

\tableofcontents


\section{Introduction}

Our comprehension of the dynamics of strongly-coupled {\it chiral} gauge theories is still largely unsatisfactory, in spite of their 
potential role in describing the physics of fundamental interactions beyond the standard model. The results of 
many years of study of these theories, by using mixture of wisdom and certain consistency conditions, the most significant among them being
't Hooft's anomaly matching constraints \cite{tHooft}, do restrict possible lists of dynamical scenarios and symmetry realization patterns \cite{Raby}\sp\cite{BK},  but they are usually far from being capable to determine the infrared physics of these systems uniquely. 

   It is customary in these discussions of the dynamics and symmetry realization in 
chiral gauge theories, to take into account the nonanomalous symmetries only.  For instance,  the 't Hooft anomaly matching requirements are normally applied exclusively on nonanomalous global symmetries.  

From the point of view of renormalization group,  assuming that the color interactions become strongly coupled towards the infrared, and that the low-energy effective degrees of freedom are not the original gluons and quarks (and similar color gauge bosons and matter fermions in general chiral theories),  the anomalies present in the underlying theory
must be reproduced for consistency either by composite massless fermions  ('t Hooft) if the symmetry remains unbroken, or, if spontaneously broken,  by massless Nambu-Goldstone (NG) bosons. In the latter case, the condition of  ``anomaly matching" is nothing but the well-known procedure for calculating the amplitudes  (such as $\pi \to \gamma\gamma$, or much more generally,
all anomalous amplitudes included  in Wess-Zumino-Witten effective action) containing the NG bosons.

A third type of the application of renormalization-group invariance of anomaly concerns certain $U(1)$ symmetry(ies), which is (are) affected by the 
topologically nontrivial gauge field configurations.  This phenomenon goes under the name of strong anomaly, and the ``$U(1)_A$ problem" and its solution in QCD
\cite{WittenU1,VenezianoU1}  is a renowned example in which the strong anomaly plays the central role. Probably because this appeared in the context of some characteristic aspects of QCD which is a vectorlike gauge theory,  such as the chiral symmetry breaking and the low-energy sigma models,  a similar question   
has not been discussed much in the context of strongly-coupled chiral gauge theories, to the best of our knowledge, 
with an exception being \cite{Vene}. 

The aim of the present note is to discuss the implication of strong anomaly in strongly-coupled chiral gauge theories.  Rather surprisingly,  the requirement that  the assumed set of infrared effective degrees of freedom (massless NG bosons and/or massless composite fermions \`a la 't Hooft) 
should be able to describe the strong-anomaly effective action in a way analogous to the famous strong-anomaly QCD effective action \cite{Rosenzweig}\sp\cite{Nath} containing certain logarithmic function,  yields a rather solid indication about which type of the infrared phase is more plausible than others.

The models we will discuss in some details are the so-called Bars-Yankielowicz (BY) and generalized Georgi-Glasow (GG)  models.
In particular,     our discussions will be set up first  by  using  two simplest classes of models.  One is  
an  $SU(N)$ gauge theory  with left-handed fermions in the reducible, complex representation, 
\be       \yng(2) \oplus   (N+4) \,{\bar   {{\yng(1)}}}\, \ee
that is, 
\beq
   \psi^{\{ij\}}\,, \quad    \eta_i^B\, , \qquad    i,j = 1,2,\ldots, N\;,\quad B =1,2,\ldots , N+4\;,
\eeq
which is the simplest of the so-called  Bars-Yankielowicz models \cite{BY}.  This model will be called ``$\psi\eta$"  model below,   for short.\footnote{In some earlier literature these fields were  denoted by
$S$ (the symmetric tensor $\psi$),  $A$ (the antisymmetric tensor $\chi$), and  ${\bar F}$  (the antifundamental,  $\eta$).}
The global symmetry group (actually the local property of the symmetry group) is 
\be   G_{\rm f}=     SU(N+4) \times U(1)_{\psi\eta}\;,     \label{beyond1}
\ee
where $U(1)_{\psi\eta}$ indicates the anomaly-free combination of $U(1)_{\psi}$ and  $U(1)_{\eta}$, associated with the two types of  matter Weyl fermions of the theory. 
Another model we consider is an $SU(N)$ gauge theory with fermions
\be       \yng(1,1) \oplus   (N-4) \,{\bar   {{\yng(1)}}}\, \ee
that is, 
\beq
   \chi^{[ij]}\,, \quad    \eta_i^B\, , \qquad    i,j = 1,2,\ldots, N\;,\quad B =1,2,\ldots , N-4\;,
\eeq
(the simplest  Georgi-Glashow model).   We will refer to it as   ``$\chi\eta$"  model below. The global symmetry group  of the $\chi\eta$ model is 
\be   G_{\rm f}=     SU(N- 4) \times U(1)_{\chi\eta}\;.   \label{beyond2}
\ee
Our interest is to understand  how these symmetries are realized in the infrared.  

  More general Bars-Yankielowicz      and 
Georgi-Glasow   models,  which are similar to  the above two models but with $p$ additional pairs of fermions in the fundamental and 
antifundamental representations, will also be considered. 

In all  these models  the conventional 't Hooft anomaly matching discussion apparently allows a confining phase, with no condensates and with full unbroken global symmetry, with some simple set of massless
 composite fermions saturating 
the anomaly matching equations.  See Appendix~\ref{conf1}  for the $\psi\eta$ model,  Appendix~\ref{conf2} for the $\chi\eta$ model and 
Appendices~\ref{conf3} and \ref{conf4} for more general classes of  BY and GG   chiral gauge theories.     

At the same time, the anomaly constraints are  also  consistent with a dynamical Higgs phase, in which the color and (part of) the flavor symmetry are  dynamically broken by certain bifermion condensates, see Appendices~\ref{Higgs1},  \ref{Higgs2},  \ref{Higgs3} and  \ref{Higgs4} 
(taken from \cite{BKL4}).   These results are mostly known from the earlier work  \cite{BY}\sp\cite{BK}
and partly completed by ourselves,  but are recorded here to make our discussion self-contained.

 There is an aspect of the conventional 't Hooft anomaly matching procedure in the dynamical Higgs phase,  which is perhaps not 
 widely appreciated. The Higgs phase of these chiral theories are, in general,  described by massless NG bosons as well as some massless fermions.   These fermions saturate the conventional 't Hooft anomaly triangles   with respect to the unbroken flavor symmetries.  The way they do is, however, quite remarkable, and in our view,  truly significant. As can be seen from Table~\ref{SimplestBis}, Table~\ref{SimplestAgain2}, and in similar 
 Tables~\ref{brsuv},   \ref{brsir},   \ref{brauv} and   \ref{brair}   for the generalized BY and GG models   (see  Appendices  \ref{Higgs1}, \ref{Higgs2}, \ref{Higgs3}, \ref{Higgs4}),  the set of fermions remaining massless in UV   and  those in the IR are  {\it identical }  as regards their quantum numbers, charges, and multiplicities. Therefore,  the matching of anomalies (in the unbroken global symmetries) is completely automatic, and is natural.   No arithmetic 
equations need to be solved to justify a solution which might look sometimes miraculous. The significance of such a solution of 't Hooft's equations 
(the dynamical Higgs phase) is its stablity \footnote{As noted in \cite{BKL2,BKL4},  in the study making use of generalized anomalies
one might try to make further gauging and study associated anomalies in the low-energy effective theory. From the identities of the sets of massless fermions in the UV and IR, it is seen that such an extra gauging would not produce any new  unmatched anomalies. 
}.

As we will find out below, the consideration of the strong anomaly appears to give us a rather clear indication that this second type of vacua - those in dynamical Higgs phase, characterized by certain bifermion condensates breaking color and part of the flavor symmetry - describe correctly the infrared dynamics of these chiral gauge theories.

This paper is organized as follows. 
In order to present the discussion systematically, we first  review  (for QCD)  and work out for the $\psi\eta$ and $\chi\eta$  theories,  the chiral Ward-Takahashi identities and identify the interpolating fields for all NG bosons, assuming  the dynamical Higgs phase for these models
(Appendices  \ref{Higgs1} and \ref{Higgs2}).  This is done in Sec.~\ref{NGB}

After this preparation,  in Sec.~\ref{sec:stronganomaly} we write the strong-anomaly low-energy  effective action for the chiral gauge theories, 
$\psi\eta$,  $\chi\eta$, as well as all other BY and GG models,  following the procedure  used for QCD.  
This discussion confirms the consistency of the dynamical Higgs phase of these models,  in agreement with the recent mixed-anomaly study  in these chiral gauge theories  \cite{BKL2,BKL4},  but it actually somewhat  strengthens the conclusion \footnote{ For instance the argument based on strong anomaly does not depend on whether $N$ is even or odd, whereas the mixed-anomaly calculation of \cite{BKL2,BKL4} were done for models with even $N$.}.

 The basic observation is that there is no way of writing the strong-anomaly effective action if  confining flavor symmetric vacua are  assumed, i.e.,  by using only the massless baryons  (Appendices \ref{conf1}, \ref{conf2}, \ref{conf3}, and \ref{conf4}).
The fermion-zero-mode (in the instanton background) counting would not work, in any of the chiral theories considered.  
 
We shall  comment briefly on some other types of theories  (other than BY and GG models) as well, and discuss the implication of the strong anomaly on their phases. 
 
Related questions regarding apparent complementarity in (only) one of the models  (the $\chi\eta$ model) and a large $N$ argument concerning $U(1)$ NG bosons in some chiral gauge theories, will be briefly commented  upon in Sec.~\ref{complementarity} and in Sec.~\ref{largeNU1}, respectively.  

In  Sec.~\ref{Summary},  our results are summarized. 
  From the analyses of Sec.~\ref{NGB} $\sim$ Sec.~\ref{largeNU1} some striking analogies and at the same time contrasts, between vectorlike theories and chiral gauge theories seem to emerge. Better stated, one perhaps learns from these discussions a more precise meaning of what is similar or what is dissimilar between the strong interaction dynamics of vectorlike and chiral gauge theories beyond certain loose use of terminologies; as a result one gets a somewhat clearer understanding of the dynamics of strongly-coupled chiral gauge theories than before.

\section{Nambu-Goldstone (NG) bosons and condensates  \label{NGB}}

	Consider any global continuous symmetry  $G_{\rm f}$  and the associated conserved current $J_{\mu}$ and charge $Q$,  the field $\phi$ (elementary or composite) which condenses and breaks  $G_{\rm f}$, and the field 
${\tilde \phi}$  which is transformed into   $\phi$  by the  $G_{\rm f}$  transformation:
\be  Q \equiv  \int d^3x    J_{0}\;, \qquad    [Q,  {\tilde \phi}] =    \phi\;, \qquad \brc \phi \ckt  \ne 0\;.   \label{WTI1}
\ee
Thus
\bea  && \lim_{q_\mu \to 0 }  i q^{\mu} \int d^4x   \, e^{- i q \cdot x} \brc  0| T\{ J_{\mu}(x) \,  {\tilde \phi}(0) \}  | 0 \ckt =    \lim_{q_\mu \to 0 }  \int d^4x   \, e^{- i q \cdot x}    \de_{\mu}   \brc  0| T\{ J_{\mu}(x) \,  {\tilde \phi}(0) \}  | 0 \ckt =  \nonumber \\
&&=   \int d^3x    \brc  0| [  J_{0}(x),   {\tilde \phi}(0)  ]  | 0 \ckt =    \brc  0| [  Q,   {\tilde \phi}(0) ]  | 0 \ckt =   \brc  0| \phi(0) | 0 \ckt  \ne 0 \;.  \label{chWT}
\eea
This Ward-Takahashi like identity  
implies that the two-point function
\be    \int d^4x   \, e^{- i q \cdot x} \brc  0| T\{ J_{\mu}(x) \,  {\tilde \phi}(0) \}  | 0 \ckt  \label{2point} 
\ee
is singular at $q^{\mu}  \to 0$.  If the $G_{\rm f}$ symmetry is broken spontaneously such a singularity is due to a massless scalar particle  
in the spectrum  -Nambu-Goldstone (NG) boson-,  a ``pion"  below, symbolically,  such that 
\be   \brc 0| J_{\mu}(q)  |  \pi \ckt =  i q_{\mu}  F_{\pi}\;, \qquad   \brc \pi | {\tilde \phi} | 0 \ckt \ne 0\;. 
\ee
Two point function  (\ref{2point})   (times $q^{\mu}$)  behaves as 
\be   \lim_{q^{\mu} \to 0} \,  q^{\mu} \cdot q_{\mu} \,  \frac{F_{\pi}   \brc \pi | {\tilde \phi} | 0 \ckt  }{q^2} \sim \const \;.  \label{WTI2}  
\ee
The constant  $F_{\pi}$  represents the amplitude for the broken current to produce the pion from the vacuum  (the pion decay constant).

\subsection{$N_{\rm f}$\,-\,flavored QCD}

  In the standard  QCD with $N_{\rm f}$  light flavors, the quarks are   
\be   \psi_L^i , \, \psi_R^i  \;,  \qquad     i=1,2,\ldots, N_{\rm f}\;. 
\ee
We take 
\be      \phi =  {\bar \psi_R} \psi_L +  {\rm{\rm  h.c.   } }  \;,  \qquad   {\tilde \phi} =   {\bar \psi_R} t^b  \psi_L -    {\rm{\rm  h.c.   } }   \;;  \qquad  J^{5, a}_{\mu} =   i  {\bar \psi_L}  {\bar \sigma }_{\mu}  t^a \psi_L - (L\leftrightarrow R)\;, \label{copiax}
\ee
where $t^a$ is the $SU(N_{\rm f})$ generators,
$a=1,2,\ldots N_{\rm f}^2-1 $.
It is believed, and confirmed by lattice simulations,   that   for sufficiently small $N_{\rm f}$, the field $\phi$  condenses, 
\be   \brc \phi \ckt=   \brc  {\bar \psi_R} \psi_L   + {\rm{\rm  h.c.   } }   \ckt \sim   - \Lambda^3\;, 
\ee
 leaving $ SU(N_{\rm f})_V\times U(1)_V$ unbroken.  The axial $SU(N_{\rm f})_A$ is spontaneously broken:
 \be   
 \brc { [  Q_5^a,  {\tilde \phi}^b ]   } \ckt
    =    -   \delta^{ab}     \brc   {\bar \psi}_R    \psi_L   +  {\bar \psi}_L    \psi_R    \ckt     =    c \,   \delta^{ab}      \,
\Lambda^3 \ne 0     \;,    \label{qcondens}
 \ee
 where $c$ is  a constant of the order of unity.
  The axial $U(1)_A$ is also spontaneously  broken, but due to the strong anomaly the associated NG boson gets mass  (the $U(1)_A$ problem: see Sec.~\ref{QCDAnomaly}).  

\subsection{NonAbelian NG bosons   in the $\psi\eta$ model  \label{NG1}} 

In the dynamical Higgs phase of the $\psi\eta$ model (Appendix~\ref{Higgs1})  the nonanomalous symmetry is broken as 
\be       SU(N) \times    SU(N+4) \times U(1)_{\psi\eta}  \xrightarrow{ \brc  \psi \eta \ckt}        SU(N)_{_{\rm cf}} \times  SU(4) \times  U(1)^{\prime}\;.  \label{symmetrypsieta}
\ee
Let us  first  
concentrate our attention to  $8N$  NG bosons associated with the $SU(N+4)$ breaking, leaving the discussion of the 
$U(1)$ NG boson and an unbroken $U(1)^{\prime}$ symmetry to the next subsection. 
An $SU(N+4)$  current   is   ($T^a$ is an $SU(N+4)$ generator)
\be    J^a_{\mu} =  i \, {\bar \eta}^{j, m}  {\bar \sigma }_{\mu}     (T^a)_{mn} \, \eta_j^{n}\;,\qquad     m,n =1,2,\ldots  N+4   
\ee
and  the charges are
\be   Q^a=  \int d^3x  \,   J^a_{0}\;. 
\ee
In particular,   consider  the broken symmetry currents  ($8 N$  components):
\be      J^a_{\mu} =  i \, {\bar \eta}^{j, m}  {\bar \sigma }_{\mu}     (T^a)_{mn} \, \eta_j^{n}\;,  
    \ee  
    with 
\be    m =1,2,\ldots  N\;, \quad   n =N+1,\ldots  N+4\;,\qquad {\rm or \,  vice \,  versa},
\ee 
with charges
\be      Q^a=  \int d^3x  \,   J^a_{0} = \int d^3x  \, {\bar \eta}^{j, m} (T^a)_{mn} \eta_j^{n}\, .       \label{Bcharge}
\ee
The associated symmetry is  broken by the condensates
\be  \brc  \psi^{\{ij\}}   \eta_j^m \ckt = \, c_{\psi\eta} \,\Lambda^3  \delta^{i   m }\;,   \qquad \quad \     i, m=1,2,\dots  N\;,     \label{psietacond}  
\ee
and there will be   $8N$  NG bosons associated with the currents,  $ J^a_{\mu}$.  
A natural choice for the pion  interpolating field  is a gauge-invariant composite,
\be      {\phi}^{\tilde a}  =   \left( \psi^{ik}  \eta_k^n    \right)^*     (T^{\tilde a})_{nm}   \, \left( \psi^{ij}  \eta_j^m  \right) \;.     \label{xpions1}
\ee
The relevant commutators are:
\be  \brc  [Q^a,   {\phi}^{\tilde a}  ]  \ckt   = -[T^a,  T^{\tilde a}]_{m n}     \brc   \left( \psi^{ik}  \eta_k^{m}   \right)^*   \, \left( \psi^{ij}  \eta_j^n  \right)    \; \ckt  \;.\label{commut}
\ee
Let us consider various $SU(2)$ subalgebras   ($\sigma^i$'s are the Pauli matrices) 
\be      T^a = \frac{\sigma^1}{2}\; \qquad  T^{\tilde a}  =  \frac{\sigma^2}{2}\;,\qquad   T^0=   \frac{\sigma^3}{2}    \label{su2}
\ee
 living in the $ 2 \times 2$ subspace
$(m, n)$,
\be    m \in  \{1,2,\ldots  N\}   \;, \quad   n \in  \{  N+1,\ldots  N+4 \}\;,  
\ee
for the broken  charge  $Q^a$  (\ref{Bcharge}) and the pion field    $  {\phi}^{\tilde a} $   (\ref{xpions1}), respectively.      The commutator (\ref{commut}) now gives
\be   \brc  [Q^a,   {\phi}^{\tilde a}  ]  \ckt   = -   \frac{1}{2}   \brc   \left( \psi^{ik}  \eta_k^{m}   \right)^*   \, \left( \psi^{ij}  \eta_j^m  \right)  \ckt 
+  \frac{1}{2}     \brc   \left( \psi^{ik}  \eta_k^{n}   \right)^*   \, \left( \psi^{ij}  \eta_j^n  \right)  \ckt \; 
\ee
({\it  no summation over $m$ or $n$ here  and below, in (\ref{analogo})})\footnote{The summation over the repeated color indices, $j$, $k$  are done,  as usual.}.
Upon condensation of color-flavor diagonal form (\ref{psietacond}), 
the above becomes  
\be     \brc  [Q^a,   {\phi^{\tilde a}} ]  \ckt   = - \frac{1}{2}   ( c_{\psi\eta}\, \Lambda^3 )^2 \ne 0\;,
\ee
implying that $Q^a$    
 generates a massless NG bosons from the vacuum, whose ``wave function" is roughly
\be     \brc 0 |     {\phi}^{\tilde a}(x)    | \pi^a(p) 
\ckt  =  f_p(x)  \;. 
\ee

Note that,   by  inserting (\ref{psietacond}),     the pion  interpolating field (\ref{xpions1})  can also be written in a simpler, gauge dependent form:
\be     {\phi}^{\tilde a}  \sim     \left( \psi^{m k}  \eta_k^n    \right)^*     (T^{\tilde a})_{nm} 
+       (T^{\tilde a})_{mn}   \, \left( \psi^{mj}  \eta_j^n    \right)  =  2 \, {\Re}  \, \{  (T^{\tilde a})_{mn}  \left( \psi^{mj}  \eta_j^n    \right)  \} \;.
 \label{analogo}   \ee
Clearly, the pairing  between the charge and the pion field  can be interchanged:
\be      T^a = \frac{\sigma^2}{2}\; \qquad  T^{\tilde a}  =  \frac{\sigma^1}{2}\;,\qquad   T^0=   \frac{\sigma^3}{2}   \label{su2bis}
\ee
(instead of (\ref{su2}));    therefore one finds  $ 4N \times 2 =   8N$  NG bosons.  

A simpler way to state the result is,  as can be seen by inspection of  (\ref{su2}),  (\ref{su2bis}), and (\ref{analogo}),  that the $8N$  NG bosons of non
 diagonal  $\frac{SU(N+4)}{SU(N)\times SU(4)}$  generators are just the real and imaginary parts of   $\psi^{mj}  \eta_j^n $ ($m \le N, \,n\ge N+1$).

\subsubsection{Colored   $\psi\eta$   NG bosons \label{colored} }

The $N^2-1$   colored NG bosons 
\bea      {\tilde \phi}^{b} &\sim &    \left( \psi^{mk}  \eta_k^n    \right)^*     (T^{b})_{nm} 
+       (T^{b})_{nm}   \, \left( \psi^{nj}  \eta_j^m    \right) \;,    \nonumber  \\
&=&    \left( \psi^{nk}  \eta_k^m    \right)^*     (T^{b})_{mn}    +       (T^{b})_{nm}   \, \left( \psi^{nj}  \eta_j^m    \right) \;,  \nonumber  \\
 &\sim &    {\Re}  \, \{   (T^{b})_{nm}   \, \left( \psi^{nj}  \eta_j^m  \right)    \}   \;,   \qquad  n \le N\;, \quad m \le  N\;,  \quad  T^b \in  su(N) \;,
\label{analogoBis}
\eea
 generated by  the condensates  (\ref{psietacond}) are absorbed by the $SU(N)$  gauge bosons  by the Englert-Brout-Higgs mechanism, so do not appear
  as physical massless particles.

\subsection{$U(1)$  NG boson(s)   in the $\psi\eta$ model  \label{NG2}}

Assuming again that we are in  the dynamical Higgs phase of the $\psi\eta$ model,   let us examine the fate of the two nonanomalous $U(1)$ symmetries, 
 $U(1)_{\psi\eta}$ and a $U(1)_{\rm f}\subset SU(N+4)$.   
One combination  is broken spontaneously, giving rise to a physical NG boson, and the other combination, $U(1)^{\prime}$, remains unbroken, as 
a manifest symmetry of the low-energy theory.

For the  $U(1)_{\psi\eta}$  symmetry,  writing the matter fermions together as
\be      \left(\begin{array}{c}\psi \\ {  \eta} \end{array}\right) =   \left(\begin{array}{c}\psi^{\{ij\}} \\\eta^1_i \\\vdots \\\eta^{N+4}_i\end{array}\right)\;,
\ee 
the $U(1)_{\psi\eta}$ current is 
\be          J^{\mu}  =   i  \big({\bar \psi} \,\, \, {\bar\eta}   \big)   \, T_{\psi\eta} \,  {\bar \sigma}^{\mu}  \left(\begin{array}{c}\psi \\\eta  \end{array}\right)\;, 
\qquad  Q=  \int d^3x \, J^0\;,
\ee
where
\be   T_{\psi\eta}  =    \left(\begin{array}{cccc}(N+4)  \, {\mathbf 1}_{  \frac{N(N+1)}{2}}   &    \\ & -(N+2)  \, {\mathbf 1}_{(N+4)\cdot N}    \\\end{array}\right)
\ee
and the $U(1)_{\psi\eta}$ charge operator  is 
\be   Q_{\psi\eta} =  \int d^3x  \,  \Big[   (N+4)  {\bar \psi}_{ij} \psi^{ij}   -    (N+2)  \sum_m  {\bar \eta}^{i}_m    \eta_i^{m}  \Big]  \;,
\ee
so
\bea   &&  [Q_{\psi\eta} ,  \psi^{k \ell} \eta_{\ell}^n ] =      \left(-(N+4)+ N+2 \right)    \psi^{k \ell} \eta_{\ell}^n   = -  2     \psi^{k \ell} \eta_{\ell}^n \;, \nonumber  \\
   &&  [Q_{\psi\eta} ,  {\bar \psi}_{k \ell} {\bar \eta}^{\ell}_n ] =      \left(N+4 -   (N+2) \right)   {\bar \psi}_{k \ell} {\bar \eta}^{\ell}_n  
    = 2    {\bar \psi}_{k \ell} {\bar \eta}^{\ell}_n\;.  
     \label{coupling1}
\eea
The condensate is (\ref{psietacond}):
\be   \brc \psi^{ij} \eta_j^m \ckt =  c_{\psi\eta}\,  \delta^{im}  \, \Lambda^3\;, \qquad  i,m \le N\;,    \label{thesame}
\ee
so
\be   \brc     [Q_{\psi\eta},  \psi^{k \ell} \eta_{\ell}^n ] \ckt   \ne 0\;,\qquad   k=n=1,2,\ldots, N\;.
\ee

On the other hand, a diagonal $SU(N+4)$  generator (it acts only on $\eta$) which mixes with 
 $U(1)_{\psi\eta}$ is
\be       T_{\rm f} =    \left(\begin{array}{ccc}\, {\mathbf 0}_{  \frac{N(N+1)}{2}}   && \\ &\ 4 \, {\mathbf 1}_{N \cdot N}  &  \\ & & -  N \, {\mathbf 1}_{4 \cdot N  }\end{array}\right)\;,
\ee
with charge operator
\be   Q_{\rm f}=  \int d^3x  \,  \Big[   4   \sum_{m=1}^N  {\bar \eta}^{i}_m    \eta_i^{m}    -  N   \sum_{m=N+1}^{N+4}   {\bar \eta}^{i}_m    \eta_i^{m}    \Big]   \;,
\ee
\be  [Q_{\rm f},    \psi^{k \ell} \eta_{\ell}^n ]  =   \begin{cases}
        \ \ \    4\,  \psi^{k \ell} \eta_{\ell}^n     & \text{for}    \qquad    n\le N  \;;  \\
        -N     \psi^{k \ell} \eta_{\ell}^n   &  \text{for}  \qquad  n > N\;.
\end{cases}   \label{couplingf}
\ee

The same condensate (\ref{thesame})  
breaks also  $Q_{\rm f} $,  as  
\be   \brc     [Q_{\rm f},  \psi^{k \ell} \eta_{\ell}^n ] \ckt   \ne 0\;,\qquad     k=n=1,2,\ldots, N\;. 
\ee
But 
the combination    $U(1)^{\prime}$ generated by
\be    Q^{\prime} =    2 Q_{\psi\eta}  +  Q_{\rm f}    
\ee
remains unbroken:
\be   \brc  [2 Q_{\psi\eta}  +  Q_{\rm f} ,     \psi^{k \ell}  \eta_{\ell}^n ] \ckt =0\;  \qquad  (\forall  k,  \forall n)\,.
\ee

Any other combination of   $Q_{\rm f} $ and   $Q_{\psi\eta} $     is spontaneously broken, 
see Eqs.~(\ref{coupling1}),  (\ref{couplingf}), so there is one physical $U(1)$ NG boson in this model.  
A natural choice for the interpolating field for this physical $U(1)$  NG boson  would be  
\be    \sum_{n,j}^N   \psi^{n j} \eta_{j}^n\   = \Tr (\psi\eta)   \propto   {\mathbf 1} +   \frac{i}{F_{\pi}^{(0)}}\,  {\phi}_0  + \ldots    \;,  \label{pionpsieta}
\ee
where  the field are appropriately normalized  and $F_{\pi}^{(0)}$  is a constant  with a mass dimension.    
This  is an analogue of $ {\bar \psi_R} \psi_L  =   {\bar u_R} u_L + {\bar d_R} d_L +\ldots$ in QCD, and also analogous to the nonAbelian NG bosons,   (\ref{analogo}).  

Unlike $ {\bar \psi_R} \psi_L $ in QCD,  (\ref{pionpsieta}) is not gauge-invariant.    However, it is not difficult to find a natural   
gauge-invariant  form for the interpolating  field for the same NG boson: it could be written as
 
\be   \det   U    \;, \qquad        U^{k \ell}  =    \psi^{k j} \eta_{j}^{\ell}    \;, \label{gaugeinvCond}
\ee
by using the first $N$ flavors, $\eta_j^a$,   $a=1,2,\ldots, N$.   Note that this  composite field is a singlet of  the surviving symmetry of the $\psi\eta$ system, 
(\ref{symmetrypsieta}).  
 This will play an important role in the discussion of the strong anomaly below.    

Expanding around the VEV,  (\ref{thesame}),  (\ref{pionpsieta}) and  (\ref{gaugeinvCond}) give the same 
physical field, 
\be  \sim   \const +     \sum_{k,j}^N   \left(\psi^{k j} \eta_{j}^k\right)^{\{q\}}\;,    \label{U1NGB}
\ee
where $\left(\psi^{k j} \eta_{j}^k\right)^{\{q\}}$  indicates the fluctuation part of the composite field,  $\psi^{k j} \eta_{j}^k$.

So far,  we considered  two  nonanomalous $U(1)$ symmetries,    $U(1)_{\psi\eta}$ and a $U_f(1)\subset SU(N+4)$, and found that  one combination remains 
a manifest symmetry,  call it  $U(1)^{\prime}$,  
  while the other   (let us indicate as $U(1)_{\rm NG}$)  gets broken and generates  
an  associated, physical  massless NG boson.

Actually,  the system possesses one more,  independent,  $U(1)$ symmetry, though anomalous. Any combination of  $U(1)_{\psi}$ and $U(1)_{\eta}$ other
 than  $U(1)_{\psi\eta}$ is anomalous  (let us call it $U(1)_{\rm an}$) hence cannot be expressed as a linear combination of  $U(1)^{\prime}$ and $U(1)_{\rm NG}$.  
 
It is perhaps useful to compare the situation here 
with the massless QCD vacuum, with quark condensate, (\ref{qcondens}). 
In the latter, there is one manifest symmetry, $U(1)_V$, and an axial symmetry $U(1)_A$ which is anomalous, and is also spontaneously broken 
 (associated with a would-be NG boson which becomes massive by the strong anomaly, as reviewed below, Sec.~\ref{QCDAnomaly}).  
 In the $\psi\eta$ model under consideration, there is one manifest symmetry, $U(1)^{\prime}$,  one spontaneously broken, 
nonanomalous, $U(1)_{NG}$, with an associated  physical massless NG boson, {\it and}  an anomalous $U(1)_{\rm an}$ symmetry, 
independent of $U(1)^{\prime}$ and $U(1)_{\rm NG}$.   There is thus one  more (counting also anomalous and/or spontaneously broken)  $U(1)$  symmetry 
in our model, as compared to the massless QCD.  

It is tempting to regard the $U(1)_{\rm an}$ symmetry as an analogue of the  $U(1)_A$  symmetry in QCD.  
However,  as it turns out,  the nature of  $U(1)_{\rm an}$  (whether or not  it is spontaneously broken),  
and related to that question, the true form of the interpolating field for the physical massless NG boson  ((\ref{pionpsieta}), (\ref{gaugeinvCond}), or something else),  
can only be found after an appropriate study of the strong anomaly effective action for the $\psi\eta$ system. This will be done in Sec.~\ref{strongpsieta}.  
   Somewhat surprisingly, it will be found that the dynamical Higgs phase of the $\psi \eta$  model is not exhaustively characterized by the condensate of $\det U$.  The proper form of the strong-anomaly effective action leads to a second,   
 bi-baryon type condensate besides  $\brc \det U \ckt$,     see  (\ref{condenseBB}). This turns out to be the missing piece of the puzzle.  The true interpolating function for the physical massless $U(1)_{\rm NG}$ boson will be found, not 
to be given by $\phi_0$ of  (\ref{pionpsieta}), but  by $\phi$ in (\ref{physical}).

\subsection{$U(1)$ symmetries in the $\chi\eta$ model    \label{U1chieta}}    

In the dynamical Higgs phase of the $\chi\eta$ model the symmetry breaking proceeds as 
\bea        SU(N) \times    SU(N-4) \times U(1)_{\chi\eta}  &\xrightarrow{ \brc  \chi \eta \ckt}&     SU(N-4)_{\rm cf} \times  SU(4)_{\rm c} \times  U(1)^{\prime}
\nonumber \\   &\longrightarrow&
  SU(N-4)_{\rm cf} \times  U(1)^{\prime}\;,    \label{symmetrychieta}
\eea
where the residual color $SU(4)_{\rm c }$,  unbroken by the condensates
\be  \brc  \chi^{ij} \eta_j^m  \ckt  = c_{\chi\eta}\, \delta^{im} \, \Lambda^3 \;, \qquad  i,m =  1,2,\ldots, N-4\;,   \label{chietacond}
\ee
evolves further towards infrared, confines, develops another condensate,
\be  \brc  \chi \chi \ckt \ne 0\;, 
\ee
and gives rise to a hidden (dark matter?) sector
invisible to the massless sector.

For the  $U(1)_{\chi\eta}$  symmetry,  writing the matter fermions together as
\be        \left(\begin{array}{c}\chi \\ {  \eta} \end{array}\right) =   \left(\begin{array}{c}\chi^{[ij]} \\\eta^1_i \\\vdots \\\eta^{N-4}_i\end{array}\right)\;,    \label{cer}
\ee 
the $U(1)_{\chi\eta}$ current is 
\be          J^{\mu}  =   i  \big({\bar \chi} \,\, \,{\bar\eta}  \big)   \, T_{\chi\eta} \,  {\bar \sigma}^{\mu}  \left(\begin{array}{c}\chi \\\eta \end{array}\right)\;,\qquad  Q=  \int d^3x \, J^0\;,
\ee
where
\be   T_{\chi\eta}  =     \left(\begin{array}{cccc}(N-4)  \, {\mathbf 1}_{  \frac{N(N-1)}{2}}   \, &    \\ & -(N-2)  \, {\mathbf 1}_{(N-4) \cdot N }    \\\end{array}\right) \ee
and the charge operator  is 
\be   Q_{\chi\eta} =  \int d^3x  \,  \Big[   (N-4)  {\bar \chi}_{ij} \chi^{ij}   -    (N-2)  \sum_m  {\bar \eta}^{i}_m    \eta_i^{m}  \Big]  \;,
\ee
so
\bea  &&   [Q_{\chi\eta} ,  \chi^{k \ell} \eta_{\ell}^n ] =      \left(-(N-4)+ N-2 \right)    \psi^{k \ell} \eta_{\ell}^n   =      2    \, \chi^{k \ell} \eta_{\ell}^n   \;, \nonumber \\
  &&  [Q_{\chi\eta} ,  {\bar \chi}_{k \ell} {\bar \eta}^{\ell}_n ] =      \left(N-4 -   (N-2) \right)   {\bar \chi}_{k \ell} {\bar \eta}^{\ell}_n   = -  2  \,   {\bar \chi}_{k \ell}\, {\bar \eta}^{\ell}_n\;.  \label{coupling1Bis}
\eea
The  condensate  (\ref{chietacond}) leads to
\be   \brc     [Q_{\chi\eta},  \chi^{k \ell} \eta_{\ell}^m ] \ckt   \ne 0\;,\qquad   k=m=1,2,\ldots, N-4\;.
\ee

On the other hand, a diagonal color  $U(1)_{\rm c}\subset  SU(N)$  generator  (which mixes with 
 $U(1)_{\psi\eta}$)  is
\be       T_{\rm c} =    \left(\begin{array}{cc}     4 \, {\mathbf 1}_{N-4}  &  \\ & -  (N-4) {\mathbf 1}_4\end{array}\right)\;,  \label{eaten}
\ee
with charge operator $Q_{\rm c}$  
\be  [Q_{\rm c},    \chi^{k \ell} \eta_{\ell}^n ]  =   \begin{cases}
        4\,  \chi^{k \ell} \eta_{\ell}^n     & \text{for}    \qquad    k   \le N-4  \;;  \\
        -(N-4)   \,  \chi^{k \ell} \eta_{\ell}^n   &  \text{for}  \qquad  k  > N-4\;.
\end{cases}   \label{coupling2Bis}
\ee
The same condensate (\ref{chietacond})  
breaks  both $U(1)_{\rm c}$ and $U(1)_{\chi\eta}$, but
the combination    $U(1)^{\prime}$ generated by
\be     2 Q_{\chi\eta}  -  Q_{\rm c}    
\ee
remains unbroken:
\be   [  2 Q_{\chi\eta}  -    Q_{\rm c},     \chi^{k \ell} \eta_{\ell}^n] =0\;.    \label{unbrokenchieta}
\ee
Unlike in the $\psi\eta$ model,  therefore,  no physical massless NG bosons  appear in the $\chi\eta$ model,  as the potential NG boson
 is eaten up by the 
$U(1)_{\rm c}\subset SU(N)$ color gauge boson by the Englert-Brout-Higgs mechanism.

As in the $\psi\eta$ model,  there is actually another $U(1)$  symmetry  (any combination of  $U(1)_{\chi}$ and  $U(1)_{\eta}$, other than 
$U(1)_{\chi\eta}$).  This $U(1)$ is also  spontaneously broken by the condensates,  (\ref{chietacond}), but the associated pseudo-NG boson gets mass by 
strong anomaly, as in the $U(1)_A$ NG boson of QCD.   The fact that  there are no physical, massless NG bosons in the $\chi\eta$ model does not
 mean that this anomalous $U(1)$ symmetry  is unimportant for the discussion of the infrared dynamics, see the next section. 
 
A closely related question is what is eventually the gauge-invariant form of the condensate,  (\ref{chietacond}).  A natural choice for the $\chi\eta$
model is    ($(\chi\eta)^{im} \equiv  \chi^{ij} \eta_j^m$)
\bea      U &=&    \epsilon_{i_1 i_2 \ldots i_N}    \epsilon_{m_1 m_2 \ldots m_{N-4}}    (\chi \eta)^{i_1 m_1}    (\chi \eta)^{i_2 m_2} \ldots 
 (\chi \eta)^{i_{N-4} m_{N-4}}    \chi^{i_{N-3}i_{N-2} }  \chi^{i_{N-1}i_{N} } \nonumber \\
&\sim&   \epsilon \, (\chi\eta)^{N-4} \chi\chi\;.   \label{naturalchieta} 
\eea

\section{Strong anomaly and effective Lagrangian\label{sec:stronganomaly}} 

In this section, we first review the well-known strong-anomaly effective action in QCD and then, following the same steps, write down the
analogous effective action for chiral gauge theories.  

\subsection{Strong anomaly, $U(1)$ problem and the $\theta$ dependence in QCD  \label{QCDAnomaly}} 

   In the discussion of  the dynamics of QCD
the consideration of the anomalous axial $U(1)_A$ symmetry has been quite important, in relation to the so-called $U(1)$ problem and its 
solution \cite{WittenU1,VenezianoU1}. 
Even though the NG boson(s)  associated with the anomalous $U(1)_A$   ($\eta, \eta^{\prime}$) \footnote{Here $\eta, \eta^{\prime}$ are the
  singlet  pseudoscalar mesons of the real world, as in the Particle Data Booklet.   The attentive reader will not confuse them with the Weyl fermion 
   in the chiral  $\psi\eta$ or $\chi\eta$ models being studied in the present work.}   
get mass by the strong interaction dynamics  ($m_{\eta} \gg m_{\pi}$,  $m_{\eta^{\prime}} \gg m_K$)
the presence of the anomalous $U(1)$ symmetry has a deep implication on the spontaneous breaking  of the nonanomalous chiral symmetries, 
\be     SU(N_{\rm f})_L \times    SU(N_{\rm f})_R  \to   SU(N_{\rm f})_V,   
\ee
which generates physical lightest NG bosons, the pions. 

Such a logical connection is best seen in the effective Lagrangian approach for QCD in the large $N$ limit.  Generalizing the standard sigma model Lagrangian to include the effect of strong anomaly,   the authors of \cite{Rosenzweig}\sp\cite{Nath}  write  
\be    L = L_0 + {\hat L}  \;,     \ee
where $L_0$ is the standard sigma model effective Lagrangian
\be  L_0=    \frac{F_{\pi}^2}{2}     \Tr   \,   \de_{\mu}  U   \de^{\mu}  U^{\dagger}     +    \Tr  M \, U  +{\rm  h.c.   }+   \ldots \;; \qquad  U \equiv  {\bar \psi}_R  \psi_L \;,
\ee
and $ {\hat L}  $  represents the strong anomaly
\be       {\hat L}  =     \frac{i}{2}  q(x)  \, \log \det  U/ U^{\dagger}    +\frac{N}{a_0 F_{\pi}^2}  q^2(x)  - \theta \, q(x)\;,\label{QCDanomeff} 
\ee   
$q(x)$ is the topological density 
\be q(x)  =  \frac{g^2}{32\pi^2}  F_{\mu\nu}^a  {\tilde F}^{a, \mu\nu}\;,      \label{topodens}
\ee
$a_0$ is a constant of the order of unity,  $F_{\pi}$ the pion decay constant, and $\theta$ is the QCD vacuum parameter.  
The $U(1)_A$ anomaly under 
\be     \Delta  S  =   2 N_{\rm f}  \alpha  \int  d^4x   \frac{g^2}{32\pi^2}  F_{\mu\nu}^a  {\tilde F}^{a, \mu\nu}\;,  \qquad \psi_L \to e^{i \alpha} \psi_L\;, \quad   \psi_R \to e^{- i \alpha} \psi_R\;,
\ee
is reproduced by the  $ \log \det  U/ U^{\dagger}$ term of the effective action.

Treating $q(x)$ as an auxiliary field, and integrating, one gets another form of the anomaly term \cite{Rosenzweig}\sp\cite{Nath}
\be   {\hat L}  =    -  \frac{F_{\pi}^2  \, a_0 }{4 N}   \Big( \theta -   \frac{i}{2}  \log \det U/U^{\dagger} \Big)^2\;. \label{analog}
\ee

It has been noted that  a multi-valued effective Lagrangian involving  $\log \det  U/ U^{\dagger}$  
is only well defined because  
\be      \brc U \ckt  \propto {\mathbf 1}    \ne 0\;:
\ee
the effective potential is defined as its expansion around its VEV \footnote{We indicated the $U(1)_A$ NG boson as $\eta$ here, as the real-world pseudoscalar mesons  $\eta$ or $\eta^{\prime}$.
The attentive reader will not confuse it with the $\eta$  field in the $\psi\eta$ model under consideration.  
 }, 
\be    U \propto  e^{  i   \tfrac{ \pi^a  t^a}{F_{\pi}}   + i    \tfrac{ \eta \, t^0}{F_{\pi}^{(0)}  }   }  =    {\mathbf 1} +    i   \frac{ \pi^a  t^a}{F_{\pi}}   + i    \frac{ \eta \,  t^0}{F_{\pi}^{(0)}  }  +\ldots 
\ee

Inverting the logics, one might actually  argue that the presence of such an effective action reproducing the strong anomaly  implies a nonvanishing condensate,
$\brc U \ckt =   \brc \bar \psi_R \psi_L \ckt \ne 0$, and hence  indirectly the spontaneous breaking of {\it  nonanomalous}   chiral symmetry,  
$ SU(N_{\rm f})_L \times  SU(N_{\rm f})_R \to   SU(N_{\rm f})_V$, affecting the low-energy physics.

\subsubsection{Veneziano\,-Yankielowicz and Affleck-Dine-Seiberg  superpotentials}

  In the context of ${\cal N}=1$  supersymmetric gauge theories, the strong-anomaly effective action is expressed  by the so-called 
Veneziano-Yankielowicz   (VY) and  Affleck-Dine-Seiberg (AffDinSei)    superpotentials 
\cite{VY}\sp\cite{AffDinSei}.  They correctly reproduce  in the infrared effective theory the effects of instantons, supersymmetric Ward-Takahashi identities,  and  the anomaly of \cite{KonishiAnom,KonishiShizuya}.  The VY  and ADS superpotentials are crucial in determinig the infrared dynamics  and phases  of the ${\cal N}=1$  supersymmetric gauge theories (see  \cite{Amati:1988ft} for a review).

\subsection{Strong anomaly and effective action   in the  $\psi \eta$  model    \label{strongpsieta}}

Unlike  $ {\bar \psi_R} \psi_L$   in QCD,  $  {\tilde \phi}=  \sum_{n,j}^N   \psi^{n j} \eta_{j}^n$    is not gauge invariant.  This is not a problem if the system is assumed to be in dynamical Higgs phase of the $\psi\eta$  model,  
in which the low-energy symmetry is
\be   SU(N)_{_{\rm cf}}  \times  SU(4) \times U(1)^{\prime}\;,     \label{consistent} 
\ee
(Appendix~\ref{Higgs1}).
A gauge invariant form of the condensate, consistent with such a symmetry is:  
\be      \det   U  \;,     \qquad   U^{k \ell}    \equiv   \psi^{k j} \eta_{j}^{\ell} \;.
\ee
In terms of this composite field, one may write the low-energy effective Lagrangian describing the strong anomaly,
\be    {\hat L}  =  \frac{i}{2}    q(x)  \log   \det   U/ U^{\dagger}\;, \qquad      q(x)  =  \frac{g^2}{32\pi^2}  F_{\mu\nu}^a  {\tilde F}^{a, \mu\nu}\;,
\label{anompsieta}
\ee 
which looks very much in analogy with the strong-anomaly effective action  of QCD, (\ref{QCDanomeff}).\footnote{$\theta$ parameter is absent  in chiral gauge theories.}   The  multivalued, logarithmic potential is  well defined, as we are in the dynamical Higgs phase,  $\brc U \ckt \propto {\mathbbm 1}$. 
As in QCD, one may actually reverse the logics and argue that  the strong anomaly effective action,  which should be present in the low-energy effective theory for faithfully representing all the symmetries of the UV theory, {\it  implies} the nonvanishing condensates, $\brc \det U \ckt \ne 0$, 
which means the global symmetry breaking as in (\ref{consistent}), i.e., the system is in Higgs phase.

 Though  (\ref{anompsieta}) appears to be a natural choice in the broken phase,  it has a defect 
 of not being invariant under the full symmetry of the underlying theory,
 \be   SU(N)\times SU(N+4) \times  U(1)_{\psi\eta}\;. \label{UVconsistent} 
 \ee
 A correct low-energy effective action should be invariant under the full symmetry,  and must describe (at least, be consistent with) the breaking 
 from (\ref{UVconsistent}) to   (\ref{consistent}).   This observation brings us back to the questions on the nature of  $U(1)$ NG bosons, raised at the end of Sec.~\ref{NG2}.
 
 
 In order to get to the right form of the strong-anomaly effective action, we start from the beginning,  
 \be   {\cal L}  = -  \frac{1}{4} F_{\mu\nu} F^{\mu\nu}  +      {\cal L}^{\rm fermions}   
 \ee
 \be     {\cal L}^{\rm fermions}   
= -i \overline{\psi}{\bar {\sigma}}^{\mu}\left(\partial +\calR_{\rmS}(a)   \right)_{\mu}  \psi\;  
- i   \overline{\eta}{\bar {\sigma}}^{\mu} \left(\partial +  \calR_{\rmF^*}(a) \right))_{\mu}   \eta\; \label{naivepsipeta}
\ee
($a$ is the $SU(N)$ gage field, and  the matrix representations appropriate for $\psi$ and $\eta$ fields are indicated in an obvious notation). 
Change the variables by 
 \be   {\cal L}  = -  \frac{1}{4} F_{\mu\nu} F^{\mu\nu}  +      {\cal L}^{\rm fermions}   +  \Tr [(\psi\eta)^* U]    +{\rm  h.c.   } 
 +   { B}\,(\psi\eta\eta)^* +{\rm h.c.}\;,
 \ee
 where  $U$  is the composite scalars of $N\times (N+4)$ color-flavor mixed matrix form,
 \be      \Tr [(\psi\eta)^* U]   \equiv    (\psi^{ij}  \eta_j^m)^*  U^{im}  \;
 \ee
and $ { B}$  are the baryons   $ B \sim \psi \eta \eta$,  
     \be     B^{mn}=    \psi^{ij}  \eta_i^m  \eta_j^n \;,
       \label{baryons00}
\ee
antisymmetric in  $m  \leftrightarrow n$.   Here we have anticipated the fact that these baryonlike composite fields, present in the Higgs phase together with the composite scalars $~\psi\eta$  (see Appendix~\ref{Higgs1}),    are also needed to write  down the 
 strong-anomaly effective action, see below. 

 Integrating over $\psi$ and $\eta$  one gets    
 \be   {\cal L}^{\rm eff}  = -  \frac{1}{4} F_{\mu\nu} F^{\mu\nu}  + \Tr  ({\cal D}  U)^{\dagger}    { \cal D} U     -i  \overline{ B}   \, { \bar {\sigma}}^{\mu}   \partial_{\mu} { B}      -  V\;.     \label{potential}   \ee
 The potential  $V$ is assumed to be such that its minimum is of the form,  (\ref{psietacond}):    
\be  \brc  U^{im}  \ckt = \, c_{\psi \eta} \,\Lambda^3  \delta^{i m }\;,   \qquad \quad \     i, m=1,2,\dots  N\;,     \label{condU}  
\ee
and among other terms, it contains the strong anomaly term,  $ {\hat L}$, 
\be   V  =   V^{(0)} +    {\hat L}\;,
\ee
which we choose as
\be     {\hat L}=    \const \,   \left[  \log   \left(  \epsilon \,  { B}  { B}  \det  U \right)    -  \log \left(  \epsilon \, { B}  { B}  \det  U \right)^{\dagger}  \right]^2\;.  \label{strong} 
\ee
The argument of the logarithm 
\be \epsilon  \, { B}  { B}  \det  U  \equiv    \epsilon^{m_1,m_2,  \ldots, m_{N+4}} \epsilon^{i_1, i_2, \ldots, i_N}
 { B}_{m_{N+1}, m_{N+2}}  { B}_{m_{N+3}, m_{N+4}}       U_{i_1 m_1}  U_{i_2 m_2} \ldots  U_{i_N m_N}\;  \label{insteadof}
\ee
is invariant under the full (nonanomalous) symmetries,
\be  SU(N)_{\rm c}\times SU(N+4)\times   U(1)_{\psi\eta}\;.
\ee
It contains $N+2$ $\psi$'s and $N+4$   $\eta$'s, the correct numbers of the fermion zeromodes in the instanton background:
in other words,  it corresponds to  a  't Hooft's instanton n-point function, e.g.,  
\be    \brc  \psi\eta\eta(x_1) \psi\eta\eta(x_2) \psi\eta(x_3) \ldots    \psi\eta(x_{N+2}) \ckt\;.
\ee

Up to now, we have assumed that the $\psi\eta$  system is in dynamical Higgs phase, characterized by  the bifermion $\psi\eta$  condensate,  (\ref{condU}). 
In fact, a crucial observation is  that there is no way of saturating the fermion zero modes,  $N_{\psi} = N+2$; $N_{\eta} = N+4$, 
by using the baryon  fields  (${ B}\sim \psi \eta \eta$) only.  This is a strong hint that such a confining phase  in which the only infrared 
degrees of freedom are  the baryons (Appendix~\ref{conf1})  cannot be the correct vacuum of the system.

 The problem of multi-valuedness of the effective action is solved by assuming that 
 \be      \brc  \epsilon^{(4)}   { B}  { B}   \ckt \ne 0\;, \qquad   \brc    \det  U  \ckt  \ne 0\;,   \label{condenseBB}
 \ee
 where 
  \be    \epsilon^{(4)}   { B}  { B}   =    \epsilon_{\ell_1  \ell_2  \ell_3  \ell_4}   { B}^{\ell_1 \ell_2}   { B}^{\ell_1 \ell_2}\;, \qquad 
 \ell_i =  N+1,\ldots, N+4\;.      \label{asabove} 
 \ee
 As  
 \be    \brc    \det  U  \ckt     \propto  {\mathbf 1}_{N\times N}\;
 \ee
 takes up all flavors up to $N$ (by using the full $SU(N+4)$ symmetry to orient the symmetry breaking direction), 
 $  { B}  { B}  $ must be made of the four remaining flavors,  as  in (\ref{asabove}). 
 These baryons were not considered in earlier studies \cite{ADS,BKL4}, but are assumed to be massless here,
and indicated as $B^{[A_2 B_2]}$  in Table~\ref{SimplestBis} (Appendix~\ref{Higgs1}).
 Note that  this appears to present us with another puzzle.  If these extra baryons were massive,
 how could such composites appear in the low-energy effective action?   Actually, there is an elegant answer.  
 The point is that  these  extra baryons do not have conventional triangle anomalies  with respect
 to the unbroken flavor symmetry,  
 \be    G_{\rm f}^{\prime}  =  SU(N)_{\rm cf}\times SU(4) \times U(1)^{\prime}  \ee
(in particular,  they have a vanishing $U(1)^{\prime}$ charge),   as can be seen in Table~\ref{SimplestBis} (Appendix~\ref{Higgs1}). We assume therefore 
that the baryons, indicated as  $B^{[A, B]}$  in  Table~\ref{SimplestBis}, are  all massless,  including those carrying  the flavor indices
($A,B= N+1,\ldots, N+4$).
 The conventional  't Hooft criterion with respect to $SU(N)_{\rm c}\times SU(4) \times U(1)^{\prime}$   
 did not require them but does not exclude them either.

We are now ready to answer the questions raised at the end of Sec.~\ref{NG2}.  We expand around the  VEV, 
\bea    && \det  U  =  \brc \det  U\ckt +  \ldots   \propto    {\mathbf 1}  +      \frac{i}{F_{\pi}^{(0)}}  \,  \phi_0   +\ldots \;;      \nonumber \\
    && \epsilon^{(4)}   { B}  { B}   =   \brc  \epsilon^{(4)}   { B}  { B} \ckt +  \ldots  \propto 
   {\mathbf 1}   +     \frac{i}{F_{\pi}^{(1)}}  \,  \phi_1 +\ldots  \;, 
\eea
where we appropriately re-normalized the fields,  $\phi_0$ is the fluctuation of  $\sum_n (\psi\eta)^{nn}$, (\ref{pionpsieta}), and     $F_{\pi}^{(0)}$ and 
 $F_{\pi}^{(1)}$  are some constants with dimension of mass. 
 
 Note that   under the nonanomalous $U(1)$ symmetries  discussed in  Sec.~\ref{NG2},  
 \be    \brc [Q_{\psi\eta},     \epsilon  \, { B}  { B}  \det  U    ] \ckt  =0\;,  \quad   \brc [Q_{f},     \epsilon  \, { B}  { B}  \det  U    ]\ckt =0\;,
 \ee
 as expected:   the fluctuation of $ \epsilon  \, { B}  { B}  \det  U   $,
 \be      {\tilde \phi} \equiv   N_{\pi}   \left[ \frac{1}{F_{\pi}^{(0)}}   \,   \phi_0 +   \frac{1}{F_{\pi}^{(1)}}  \,  \phi_1 \right] \;, \qquad   N_{\pi} =  \frac  { F_{\pi}^{(0)} F_{\pi}^{(1)}}{ \sqrt{\big(F_{\pi}^{(0)}\big)^2 +  \big(F_{\pi}^{(1)}\big)^2 }}  \;, 
 \ee
   therefore  does not  represent the physical NG boson.    Indeed,  the strong-anomaly effective action  (\ref{strong}) gives a quadratic mass term for  ${\tilde \phi}$.
   
   Vice versa,   an orthogonal combination   
\be      {\phi} \equiv   N_{\pi}   \left[ \frac{1}{F_{\pi}^{(1)}}   \phi_0   -    \frac{1}{F_{\pi}^{(0)}}  \phi_1 \right] \;,  \label{physical}  \ee 
 does not get mass from  (\ref{strong}).     It is this field that represents  (i.e., is the interpolating field of) the physical  $U(1)_{NG}$ NG boson of the $\psi\eta$  model, not the na\"{i}ve expectation  $ \phi_0$,   discussed in Sec.~\ref{NG2}. 
   The presence of two condensates, $  \brc    \det  U  \ckt $ and 
 $  \brc  \epsilon^{(4)}   { B}  { B}   \ckt,$ is thus the key to answer the questions brought up at the end of  Sec.~\ref{NG2}.  In particular,    {\it any} anomalous  combination of  $U_{\psi}(1)$ and   $U_{\eta}(1)$   is spontaneously broken also, and  $U(1)_{an}$  can indeed be regarded as a good analogue of the axial  $U(1)_A$ symmetry in QCD.

 Let us check that everything fits together. 
 We note that $  \brc  \epsilon^{(4)}   { B}  { B}   \ckt \ne 0\;$   but   $\brc    { B}   \ckt =0$.     On the contrary, 
 $  \brc    \det  U  \ckt     \ne 0  $ implies  the condensates   $ \brc U  \ckt     \propto  {\mathbf 1}_{N\times N}$, as 
 $U \sim \psi\eta$  is  a {\it scalar} composite.  Furthermore,  the condensate $  \brc  \epsilon^{(4)}   { B}  { B}   \ckt \ne 0\;$   is  a singlet of the unbroken symmetry  ${G_{\rm f}}^{\prime}$:  it
  does not  modify the symmetry breaking pattern in the Higgs phase of the $\psi\eta$  model,  determined by the color-flavor locked bifermion condensate,
  $\brc \psi\eta \ckt \propto {\mathbf 1}$, (\ref{consistent}).    Also,   there are four-baryon couplings from (\ref{strong}),  and due to Nambu-Jona-Lasinio mechanism,  ${ B}\sim B^{[A_2 B_2]}$ acquires mass: they disappear from the massless spectrum, leaving the massless spectrum of the $\psi\eta$ model considered earlier \cite{ADS,BKL4}.  It might appear that this effectively brings us back to the simple-minded approach to the strong anomaly sketched at the beginning of this section, but 
  as we saw above,  a more careful treatment was really needed.

The full understanding of the problem would require clarification of the mechanism for the condensation of the baryon pair   together with that of $\det U$,  (\ref{condenseBB}).  The absolute values of the condensates $BB$ and $\det U =  \det (\psi \eta)$ are flat directions of the potential (\ref{strong}). The instanton-induced potential is  thus  insufficient to determine them in itself.
Even though the condensation of  $\det U$  is analogous to the quark condensates in QCD, and can be understood as due to the strong interactions among $\psi$'s and $\eta$'s with the color gauge bosons,  the condensation of the color-singlet baryons is less obvious. Being the components of the baryon $B \sim \psi\eta\eta$ charged under color, it is natural to expect residual dipole-like interactions between the $B$'s. The flat directions of the instanton potential (\ref{strong}) will be lifted, in some way, by quantum corrections.
Such an information is implicitly in the (unkown) potential $V$  in (\ref{potential}), but 
certainly a more in-depth study (as in the Coleman-Weinberg effective action)
is needed to understand the mechanism for the condensation of $BB$,  and the relation with the $\det U$ condensate.

\subsection{Strong anomaly effective action in the generalized BY models \label{BY} }  

As the solution given above on the  $\psi\eta$ model is remarkably subtle, one might wonder whether a similar mechanism is at work in the
so-called general Bars-Yankielowicz model,  
an $SU(N)$ gauge theory with Weyl fermions 
\beq
   \psi^{ij}\,, \quad    \eta_i^A\, \,, \quad    \xi^{i,a}  
\eeq
in the direct-sum  representation
\be       \yng(2) \oplus   (N+4+p) \,{\bar   {{\yng(1)}}}\;\oplus   p \,{   {{\yng(1)}}}\; .
\ee
Without repeating the analysis we recall simply \cite{BKL4} that a  chirally symmetric confining vacuum, with massless baryons
    \be    
 {({ B}_{1})}^{[AB]}=    \psi^{ij}   \eta_i^{A}  \eta_j^{B}\;,
\qquad {({ B}_{2})}^{a}_{A}=    \bar{\psi}_{ij}  \bar{\eta}^{i}_{A}  \xi^{j,a} \;,
\qquad {({ B}_{3})}_{\{ab\}}=    \psi^{ij}  \bxi_{i,a}  \bxi_{j,b}  \;,
\label{baryons10}
\ee
(the first is anti-symmetric in $A \leftrightarrow B$ and the third is  symmetric in $a \leftrightarrow b$), saturating all conventional 't Hooft anomaly triangles
(see Appendix~\ref{conf3})  
cannot be the correct vacuum of the system.    A  $({\mathbbm Z}_2)_F -  [{\mathbbm Z}_2]^2$ mixed anomaly, present in the UV theory, is absent in the IR.  

A dynamical Higgs phase (see Appendix~\ref{Higgs3})   with condensate formation,
\bea
&&  \brc  U^{iB}  \ckt= \brc  \psi^{ij}   \eta_i^B \ckt =\,   c_{\psi\eta} \,  \Lambda^3   \delta^{j B}\ne 0\;,   \qquad \ \   j,B=1,\dots,  N\;,     \nonumber \\
&&   \brc  V^{a A}  \ckt=    \brc  \xi^{i,a}   \eta_i^A \ckt =\,   c_{\eta\xi} \,  \Lambda^3   \delta^{N+4+a,  A}\ne 0\;,   \quad  a = 1,\dots, p\;,  \quad  A=N+5,\dots, N+ 4+  p \;, \nonumber \\     \label{BYvevs}
 \eea
 and  with symmetry breaking 
\bea
&& SU(N)_{\rmc}  \times   SU(N+4+p)_{\eta}  \times  SU(p)_{\xi}  \times  U(1)_{\psi\eta}\times  U(1)_{\psi\xi} \nn \\
&&  \xrightarrow{\brc  \xi   \eta \ckt , \brc  \psi \eta \ckt}      SU(N)_{{\rm cf}_{\eta}}  \times   SU(4)_{\eta}  \times  SU(p)_{\eta\xi}  \times  U(1)_{\psi \eta}^{\prime} \times  U(1)_{\psi \xi}^{\prime} \;
\label{symbres}
\eea
turns out  instead to be consistent \cite{BKL4}.

A strong anomaly effective action for these theories can be constructed in a way analogous to  the $\psi\eta$  model. 
Instead of  (\ref{insteadof}),  one has now
\bea   &&    \epsilon \  {{ B}_1}  {{ B}_1}  \det  U   \det V  
 \equiv    \epsilon^{m_1,m_2,  \ldots, m_{N+4+p}} \epsilon^{i_1, i_2, \ldots, i_N}  \epsilon^{k_1, k_2, \ldots, k_p}   \times   \nonumber \\
&& \qquad   \times  \,    { B}_1^{[m_{N+1}, m_{N+2}]}  { B}_1^{[m_{N+3}, m_{N+4}]}       U^{i_1 m_1}  U^{i_2 m_2} \ldots  U^{i_N m_N}
V^{m_{N+5}    k_1}     \ldots  V^{m_{N+4+p}  k_p}  \;, \nonumber \\
  \label{insteadofBis}
\eea
where  $B_1$  are the baryons  $\sim \psi\eta \eta$ defined in (\ref{baryons10}).  
The rest of the analysis closely follows that of the $\psi\eta$ model discussed in Sec.~\ref{strongpsieta}.
We shall not pursue further the details of  the analysis here,  except for  noting that the strong anomaly effective action with such a logarithm,  
is perfectly consistent with (implies?) the condensates, (\ref{BYvevs}), together with   
$\brc  {B}_{1}   {B}_{1} \ckt  \ne 0$, where   $ {B}_{1}  $  are the  baryons defined in   (\ref{baryons10}), (\ref{specialbaryons}).  Writing extensively,  
\bea     &&  \brc  {B}_{1}   {B}_{1} \ckt    =  \brc   \epsilon_{C_1 C_2 C_3 C_4}    (\psi^{ij} \eta_{i}^{C_1}  \eta_{j}^{C_2})      (\psi^{k\ell} \eta_{k}^{C_3}  \eta_{\ell}^{C_4})  \ckt    \ne 0\;,    \nn \\
 && C_1 \sim  C_4=N+1, \dots, N+4 \;:
\eea
i.e., the system is in dynamical Higgs phase,  described in Appendix~\ref{Higgs3}.  

On the contrary,  it is clearly not possible to write the strong-anomaly effective action  with logarithmic argument  (\ref{insteadofBis}),  
in terms of  massless composite fermions (\ref{baryons10}).  

\subsection{Strong anomaly in the  $\chi\eta$  model   \label{stronganomchieta}}

Any combination of  $U(1)_{\chi}$ and  $U(1)_{\eta}$ other than $U(1)_{\chi\eta}$  (see Table~\ref{SimplestAgain2}) suffers from the strong anomaly.  
It means that the low-energy effective action should contain a term analogous to (\ref{QCDanomeff}) for QCD or  (\ref{anompsieta}) for the $\psi\eta$
model.   The natural choice of a gauge-invariant condensate  (\ref{naturalchieta}) suggests an effective action of the form for the $\chi\eta$ model: 
\be         \frac{i}{2}    q(x)      \log      (\chi\eta)^{N-4} \chi\chi  +{\rm h.c.}\;,       \label{stronganomalychieta}    \ee    
($q(x)$ is the topological density defined in (\ref{topodens}))  where 
\bea    && (\chi\eta)^{N-4} \chi\chi  \equiv     \epsilon_{i_1 i_2 \ldots i_N}   \epsilon_{m_1 m_2 \ldots m_{N-4}}  \,  (\chi \eta)^{i_1 m_1}    (\chi \eta)^{i_2 m_2} \ldots  (\chi \eta)^{i_{N-4} m_{N-4}}    \chi^{i_{N-3}i_{N-2} }  \chi^{i_{N-1}i_{N} }\;. \nn \\  \label{zeromodes}
  \eea
The argument of the logarithmic function taken here reflects the correct number of  the fermion zeromodes in the instanton background 
($N_{\chi}= N-2$ and $N_{\eta}= N-4$);  the epsilon tensors 
take care of the invariance under the full (nonanomalous) symmetry of the $\chi\eta$   system,
\be 
 SU(N)_{\rm c} \times  SU(N- 4) \times U(1)_{\chi\eta}\;.
\ee
This anomaly effective action agrees with  the one  proposed  by Veneziano \cite{Vene} for the  special case of  $SU(5)$ $\chi\eta$  model,  
and generalizes it  to all $SU(N)$ $\chi \eta$ models.  

An important observation we share with \cite{Vene}  is that this strong anomaly effective action,  which should be present in the low-energy theory to reproduce correctly the (anomalous and nonanomalous) symmetries of the UV theory, implies nonvanishing condensates, 
\be     \brc \chi\eta \ckt\ne 0\;, \qquad  \brc \chi\chi \ckt  \ne  0\;,     
\ee         
i.e.,  the dynamical Higgs phase, Appendix~\ref{Higgs2}.

Another important observation here is  that there is no way of writing the strong anomaly effective action (\ref{stronganomalychieta})  in terms of the ``baryons",   $B \sim \chi \eta \eta$, of the assumed  confining, chirally symmetric phase (Appendix~\ref{conf2}).  No combination of the baryons can saturate the correct number of the fermion zeromodes,  {\it cfr}   (\ref{zeromodes}).

Even though, contrary to the $\psi\eta$ model,  the $\chi\eta$  system has no physical $U(1)$ NG boson  (it is eaten by a color $SU(N)$ gauge boson),  
the counting of the broken and unbroken $U(1)$ symmetries  is basically similar in the two models.  Of the two nonanomalous 
 symmetries ($U(1)_{\rm c}$ and $U(1)_{\chi\eta}$), a combination remains a manifest symmetry, and the other becomes the longitudinal part of the 
$T_{\rm c}$   gauge boson.  Still another, anomalous,  $U(1)$ symmetry exists, any combination of $U(1)_{\chi}$ and  $U(1)_{\eta}$ other than  
$U(1)_{\chi\eta}$.  This symmetry is also spontaneously broken hence must be associated with a NG boson, though it will get mass by the strong anomaly. 

By expanding  the composite  $ \chi\eta$ and $\chi\chi$  fields around their VEV's,   
\bea    && (\det U)^{\prime} =  \brc   (\det U)^{\prime}  \ckt +  \ldots   \propto      {\mathbf 1}  +    \frac{i}{F_{\pi}^{(0)}}  \, \phi_0^{\prime}    +\ldots \;,    \nonumber \\
    && \chi \chi   =   \brc  \chi \chi  \ckt +  \ldots   \propto
   {\mathbf 1}   +   \frac{i}{F_{\pi}^{(1)}}    \,   \phi_1^{\prime}   +\ldots  \;, 
\eea
where $ (\det U)^{\prime} $  is defined in the $N-4$ dimensional color-flavor mixed space, and  
\be   \chi\chi  =      \epsilon_{i_1 i_2 i_3 i_4} \chi^{i_1 i_2} \chi^{i_3 i_4}\;, \qquad    N-3  \le  i_j  \le  N\;. 
\ee
Now the strong-anomaly effective action  (\ref{stronganomalychieta})    gives mass to  
 \be      {\tilde \phi}^\prime    \equiv   N_{\pi}   \left[ \frac{1}{F_{\pi}^{(0)}}   \,   \phi_0^{\prime}  +   \frac{1}{F_{\pi}^{(1)}}  \,  \phi_1^{\prime}  \right] \;, \qquad   N_{\pi} =  \frac  { F_{\pi}^{(0)} F_{\pi}^{(1)}}{
 \sqrt{\big(F_{\pi}^{(0)}\big)^2 + \big(F_{\pi}^{(1)}\big)^2 }}  \;, 
 \ee
 whereas an orthogonal combination 
\be      {\phi}^\prime    \equiv   N_{\pi}   \left[ \frac{1}{F_{\pi}^{(1)}}     \phi_0^{\prime}  -    \frac{1}{F_{\pi}^{(0)}}  \phi_1^{\prime}  \right] \; \ee
remains massless:  it is this potential NG boson which is absorbed by the color  $T_{\rm c}$  gauge boson.

\subsection{Strong anomaly in the  generalized  GG  models   \label{stronganomGG}}

The structure of the strong anomaly action in the $\chi\eta$  turned out to be markedly simpler than that in the $\psi\eta$ model. The argument of the logarithm 
is made of composite scalar fields only.   One might wonder if such a simple description is valid also for more general  Georgi Glashow models \cite{BKL4}, 
an $SU(N)$ gauge theory with Weyl fermions 
\beq
   \chi^{[ij]}\,, \quad    \eta_i^A\, \,, \quad    \xi^{i,a}  
\eeq
in the direct-sum  representation
\be       \yng(1,1) \oplus   (N- 4+p) \,{\bar   {{\yng(1)}}}\;\oplus   p \,{   {{\yng(1)}}}\; .
\ee
It  turns out that  the  construction  (\ref{stronganomalychieta})   straightforwardly  generalizes  to  
\be         \frac{i}{2}    q(x)      \log  \epsilon  \chi\chi    \det \,  (\chi\eta) \det  (\xi\eta)   +{\rm h.c.}\;,       \label{stronganomalyGG}    \ee    
  where we used a shorthand notation
\bea   &&  \epsilon  \chi\chi    \det \,  (\chi\eta) \det  (\xi\eta)  =     \epsilon_{i_1 i_2 \ldots i_N}  \,   \epsilon_{k_1 k_2 \ldots k_p} \,  \epsilon_{m_1 m_2 \ldots m_{N-4+p} }  \nonumber \\
&&\qquad \ \   \times \,
 (\chi \eta)^{i_1 m_1}    (\chi \eta)^{i_2 m_2} \ldots  (\chi \eta)^{i_{N-4} m_{N-4}}      \chi^{i_{N-3}i_{N-2} }  \chi^{i_{N-1}i_{N} }  
  (\xi \eta)^{m_{N-3} k_{1}}    \ldots   (\xi \eta)^{m_{N-4+p} k_{p}}\;.\nonumber \\
  \label{GGAn}
\eea
Note that this is a singlet of the full symmetry group of the model.  The strong anomaly action (\ref{stronganomalyGG})  implies the condensates 
\bea
&&\, \brc  \chi^{ij}   \eta_i^A \ckt  =c_{\chi\eta} \,  \Lambda^3   \delta^{j A}\ne 0\;,   \qquad  \qquad \   j=1,\dots,  N-4\;,   \quad  A=1, \dots , N -4 \;,  \nonumber \\
&&  \brc  \xi^{i,a}   \eta_i^B \ckt =c_{\xi\eta} \,  \Lambda^3   \delta^{N-4+a, B}\ne 0\;,   \qquad  a=1,\dots,  p \;,   \quad B=N-3,\dots,N-4+  p \;,\nonumber \\
&&  \eea
{\it  and} 
\be 
\brc       \chi^{j_1 j_2}  \chi^{j_3  j_4}   \ckt      =  c_{\chi\chi} \,  \epsilon^{j_1 j_2 j_3 j_4} \Lambda^3  \ne 0\;, \qquad   j_1,\ldots, j_4 =  N-3,   \ldots, N\;.
\ee
These are {\it precisely}  the set of  condensates expected to occur  in the  Higgs phase of the  GG models,  Appendix~\ref{Higgs4} \cite{BKL4}.

Vice versa,  in the confining vacuum with unbroken global symmetry,  Appendix~\ref{conf4}, there is no way  the baryons
(\ref{baryons1}) saturate all the fermion zeromodes, as in (\ref{GGAn}).

\subsection{$SU(6)$ model with a single  fermion in  a self-adjoint 
representation   \label{sec:Yama}}  

After  the studies of BY and GG  models  above,   it may be of some interest to see whether our argument  based on the strong anomaly 
may give some relevant information  in  other types of models.   Let us discuss just a few examples.  The first is an  $SU(6)$ model 
with  a {\it single} left-handed fermion in the representation,  
\be   {\underline{20}} \, = \, \yng(1,1,1)\,, 
\ee
studied in \cite{Yamaguchi,BKL1}.  The gauging of the 1-form  $ {\mathbbm Z}_3^{C}$  symmetry  gives rise to a new anomaly of the  
nonanomalous  ${\mathbbm Z}_6^{\psi}\,$   symmetry,  
\be
{\mathbbm Z}_6^{\psi} \longrightarrow {\mathbbm Z}_2^{\psi} \;, 
\label{reduced}
\ee
 leading to a three-fold vacuum degeneracy\cite{Yamaguchi,BKL1}.  The interpretation and the details of such a breaking depends on which kind  
of condensates are formed in the infrared. As (in this particular model), a scalar bifermion composite cannot be a gauge singlet,  the author of \cite{Yamaguchi} 
suggested a four-fermion condensate
   \be  \langle  \psi \psi   \psi \psi   \rangle    \sim \Lambda^6  \ne 0 \;,    \qquad   \langle  \psi \psi  \rangle =  0 \;,     \label{thefour}
 \ee 
 whereas the authors of \cite{BKL1}  proposed the gauge symmetry breaking condensate
  \be  \langle  \psi \psi  \rangle    \sim \Lambda^3  \ne 0    \;,    \label{thefact}
\ee
with $\psi\psi$  in the adjoint representation of $SU(6)$.   Both scenarios turn out to be consistent with the discrete symmetry breaking, (\ref{reduced}), and hence agree on a three-fold vacuum degeneracy, but the infrared physics associated with these assumptions are quite different, and one wonders which of the
assumptions describes the system in the infrared.     

The strong anomaly effective action  in this model  has the form, 
\be         \frac{i}{2}    q(x)      \log      \psi \psi  \psi \psi  \psi \psi  +{\rm h.c.}\;.        \label{Yamaguchi}    \ee    
Requiring that the argument of the log acquires  a nonvanishing VEV,   the assumption of four-fermion condensate  (\ref{thefour}) appears to be somewhat unnatural, whereas the  bifermion condensates (\ref{thefact})  looks perfectly consistent,  with
\be    \langle   \psi \psi  \psi \psi  \psi \psi   \rangle  \sim      \langle  \psi \psi  \rangle^i_j  \langle  \psi \psi  \rangle^j_k  \langle  \psi \psi  \rangle^k_i  \ne 0\;, 
\ee
where the color indices are briefly  restored.

\subsection{Adjoint QCD with $N_{\rm c}=N_{\rm f}=2$ }  

Another interesting model is the adjoint QCD, widely studied in the literature. Let us however consider a particular case,  $N_{\rm c}=N_{\rm f}=2$. The fermions are  
two  Weyl fermions $\lambda^i$, $i=1,2$,   both in the adjoint of $SU(2)$.  The conventional thinking assumed that a gauge invariant bifermion condensate
\be  \brc  \lambda \lambda \ckt  \ne 0   \label{Conv}  
\ee
forms, breaking the flavor symmetry as  $SU(2)_{\rm f} \to SO(2)_{\rm f}$, leading to $2$ NG bosons, and reducing the discrete ${\mathbbm Z}_8$ symmetry  to  ${\mathbbm Z}_2$ 
resulting four degenerate vacua.  

A special interest in this model was raised by the work by Anber and Poppitz \cite{AnbPop1},  which postulates that  the system develops a condensates,
\be     \brc  \lambda \lambda \lambda \lambda  \ckt  \ne 0\;, \qquad   \brc  \lambda \lambda \ckt = 0\;,    \label{AnbPo} 
\ee
breaking the discrete symmetry  as  ${\mathbbm Z}_8$ symmetry  to  ${\mathbbm Z}_4$, hence predicting a doubly degenerate vacua.  The flavor $SU(2)_{\rm f}$
remains unbroken; massless baryons
\be    \sim  \lambda\lambda\lambda  
\ee
(which is necessarily a doublet of $SU(2)_{\rm f}$)  saturates Witten's $SU(2)$ anomaly. 

The generalized, mixed  anomaly study  in this model  predicts \cite{AnbPop1, BKL1} the discrete symmetry breaking   ${\mathbbm Z}_8 \to {\mathbbm Z}_4$:  such a result is consistent with both of the dynamical scenarios mentioned above, which are markedly different  from physics point of view.   

Does our strong-anomaly argument tell anything significant?     The analogue of  the strong anomaly effective action,   such a (\ref{QCDanomeff},   (\ref{stronganomalychieta})  and  (\ref{Yamaguchi}),  is   in this case, 
 \be         \frac{i}{2}    q(x)      \log    \lambda \lambda  \ldots  \lambda  +{\rm h.c.}\;,          \label{hehe}
 \ee
with eight $\lambda$'s  inside the argument of the logarithmic function.  In this model,  in contrast to what we saw in the preceding model, Sec.~\ref{sec:Yama},  our algorithm  seems to be consistent with either of the dynamical possibilities,    
 (\ref{Conv}), or  (\ref{AnbPo}).   

It would be interesting if a lattice study could give a definitive answer.    

As a side comment, in the case with $N_{\rm f}=1$, arbitrary $N_{\rm c}$,  the adjoint QCD becomes ${\cal N}=1$ supersymmetric Yang-Mills, and  (\ref{hehe}) 
with  $2N_{\rm f} N_{\rm c} =  2 N_{\rm c}$ ~$\lambda$'s,  reduces precisely to the Veneziano-Yankielowicz \cite{VY} effective potential.  And in that case, the assumption of the bifermion condensate, $\brc  \lambda \lambda \ckt\ne 0$,  the breaking of the discrete symmetry  ${\mathbbm Z}_{2 N_{\rm c}} \to {\mathbbm Z}_2$, and the resulting $N_{\rm c}$ fold degeneracy of the vacua (equal to Witten's index),  are  by now generally accepted as the correct answer for these systems \footnote{Lattice studies \cite{Giedt}-\cite{Piemonte} also seem to confirm this. We thank Stefano Piemonte for bringing these references to our attention. }.

\section{Complementarity in the $\chi\eta$ model?  \label{complementarity} }

In the $\chi \eta$  model  the symmetry of the massless sector  in the dynamical Higgs phase  (\ref{symmetrychieta})  (Appendix~\ref{Higgs2})  happens to coincide \footnote{As noted in \cite{BKL4} this occurs only for  the $\chi\eta$ model.  
In all other  generalized  Bars-Yankielowicz and 
Georgi-Glasow models, the dynamical Higgs phase has global symmetry distinct  from that in the confining, no-condensate phase, in spite of the fact that  one of the bifermion condensate is in the fundamental representation of the gauge group.   See Appendices~\ref{conf1}-\ref{Higgs4}.
}  with that of the UV theory hence with that of the symmetric, confining phase  (Appendix~\ref{conf2}).  This fact might lead  some, based on  the so-called  complementarity picture \cite{Fradkin},  to think 
that in the $\chi\eta$ model  these two phases 
are actually one and the same. 

The following discussion however shows that the coincidence of the global symmetries 
in the two candidate vacua in the $\chi\eta$ model is an accidental one. 
There are, indeed, clear indications that they are physically distinct \cite{ADS}.  Of course it would be difficult to think that a  phase in which 
\be   \brc   \chi \eta \ckt \propto  \Lambda^3  \ne 0 \;,  \qquad     \brc   \chi \chi \ckt \propto  \Lambda^{\prime\, 3}  \ne 0      \label{dynamical}
\ee
(with no adjustable constants in front, and $\Lambda  \sim \Lambda^{\prime}$ is the dynamical mass scale generated by the  strong $SU(N)$  gauge interactions.), 
and another, without any condensates, can be the same phase.   There are no coupling constants which can be varied continuously so that the two possible ``phases"  
(one with (\ref{dynamical}), one with no  condensates) are connected without phase transition. 

  Another relevant issue could be the fact  that 
 in the dynamical Higgs phase (Appendix~\ref{Higgs2})
there appear  $(N-4)^2-1$  massive gauge bosons of degenerate mass,  by the Higgs mechanism together with the remnant  
color-flavor diagonal $SU(N-4)$ global symmetry.  This is a clear prediction which can be tested by lattice simulations. 
Even though  it is true that these massive bosons can be re-expressed as gauge-invariant composites,  there is no particular reason why precisely these massive bosons and not others 
should appear at the mass scale $\sim \Lambda$, if the system is in confining, symmetric  phase of Appendix~\ref{conf2}.

   Also,  to understand the infrared dynamics, it is indispensable to take into account of the effects of strong anomaly correctly.  Its natural form is  (\ref{stronganomalychieta}), as discussed in the previous section, by taking into account the fermion zeromodes in an instanton background. 
 The {\it solution}  of the  effective equation of motion would  then lead to a massive NG boson,  but as in QCD and in the $\psi\eta$ model, the very solution implies  nonvanishing vacuum condensate, (\ref{dynamical}).  It means that  the system is in  dynamical Higgs phase.  
 
 Crucially,   as already  noted at the end of Sec.~\ref{stronganomchieta},  there is {\it  no way} of writing the strong-anomaly effective action in terms of the massless composite ``baryons" ($\sim \chi\eta \eta$):  the zero-mode counting  ($N_{\chi}= N-2  >  N_{\eta}= N-4$) cannot work.  This is a clear sign that the 
 ``confining phase"  (Appndix~\ref{conf2})  can  neither be equivalent to the Higgs phase  (Appendix~\ref{Higgs2}) nor is one of  consistent vacuum phases of the $\chi\eta$ model.

  Of course, one of the strongest arguments which discriminates the two candidate vacua   comes from the mixed anomalies
 and the consequent, generalized-anomaly-matching analysis \cite{BKL2,BKL4}   (see also \cite{Tong}).       
     The presence of a mixed anomaly  of the type $({\mathbbm Z}_2)_F -  [{\mathbbm Z}_N]^2 $   in the UV theory and the absence of this anomaly in the ``symmetric confining phase",  show that such a phase cannot be the correct vacuum of the theory \cite{BKL2,BKL4}.    As in the $\psi\eta$ model,  
  in the dynamical Higgs phase characterized by the bifermion condensates (\ref{dynamical}), the color-flavor locked ${\mathbbm Z}_N$  symmetry 
  is spontaneousy broken by the condensates, as  the $U(1)_{\chi\eta}$ symmetry used to form  ${\mathbbm Z}_N \subset  U(1)_{\chi\eta} \times ({\mathbbm Z}_2)_F$ is spontaneously broken.
  
To summarize:   
both the mixed-anomaly analysis \cite{BKL2,BKL4}  and the consideration of the strong-anomaly effective action Sec.~\ref{stronganomchieta},
lead to the same conclusion that one of the candidate phase (dynamical Higgs phase) is consistent whereas the other (confining, flavor symmetric phase)
is inconsistent,  but {\it  not} that these two phases are inequivalent, perhaps separated by a phase transition of some sort.

\section{Large $N$ planar dominance and $U(1)$ NG bosons in  chiral gauge theories  \label{largeNU1}   }

Large $N$ counting and  planar Feynman diagram dominance are perturbative concepts and are not adequate for describing condensates.
Nevertheless, an argument was presented  some time ago \cite{Eichten:1985fs} which states that in some chiral gauge theories such as  $\psi\eta$ or $\chi\eta$ models,  a $U(1)$ symmetry cannot be spontaneously broken,  if the system is assumed to confine. 
Such a statement, if taken without due care, might mislead the reader to conclude that, e.g., in the $\psi\eta$ model,  the only possible phase is the chirally symmetric phase, with no condensates and with massless composite fermions  (Appendix~\ref{conf1}).\footnote{Let us remember that  the notation in  \cite{Eichten:1985fs}  and in the present work  are simply related by    $S  = \psi$;  ${\bar F} =  \eta$.}

To be scrupulous, the authors of \cite{Eichten:1985fs}   {\it   do not claim} that a dynamical Higgs phase
(with gauge dependent bifermion condensate)  is impossible in the $\psi\eta$ model, or in general 
chiral gauge theories.    Indeed one of the two main conclusions of  \cite{Eichten:1985fs}  is that 
 a dynamical Higgs phase is likely to be the correct answer for some chiral gauge theories  (e.g., the model in Sec.~5 of \cite{Eichten:1985fs}). 
Their claim  concerning the $\psi\eta$ model (Sec.~3 of \cite{Eichten:1985fs}) is that {\it  if the $\psi\eta$ system  is in an exact confinement phase, then a two point function of
a broken $U(1)$ current  $  \brc 0| T \{ J_{\mu}(x)  J_{\nu}(0)\}|0 \ckt $  cannot have, in the leading $N$ approximation,  an intermediate  state 
having the  quantum numbers of the  $U(1)$  NG boson, therefore a spontaneous $U(1)$ breaking is not possible.   }

Nevertheless, there seem to be a number of issues in this  argument.  First of all,  no definition of  ``exact confinement" is
given in   \cite{Eichten:1985fs}.
We noted already that there is no global center symmetry which can be used with the Polyakov loop, to set a criterion for discriminating confinement/Higgs phases.  
The authors of  \cite{Eichten:1985fs} describe the ``confinement phase"  as a phase in which all  observable states are gauge invariant, but this is also problematic, as in a local gauge theory gauge noninvariant (gauge-dependent) states can always be re-expressed in a (though more complicated)  gauge-invariant representation.  See (\ref{pionpsieta})  versus (\ref{gaugeinvCond}), for instance.

If the distinction  (``exact confinement" or not) 
is whether or not {\it a gauge-dependent condensate} forms, this cannot be used as a criterion either:  as shown in Sec. \ref{NG2},   the  gauge-dependent $\psi\eta$ condensate can be re-expressed as a gauge invariant multifermion condensate,   (\ref{gaugeinvCond}).  

If ``exact confinement" means, in the context of the $\psi\eta$ model, that no condensates  $\brc \psi\eta \ckt  \ne 0 $ 
form,  then no  $U(1)$ symmetry is broken spontaneously, hence no NG bosons would appear in the system anyway, 
independently of the $\frac{1}{N}$ counting.  The observation  \cite{Eichten:1985fs}  that in the large $N$ limit, assuming the planar diagram dominance,  the intermediate states are all made of  pairs of outgoing and incoming  $\psi$'s and $\eta$'s, hence are all neutral with respect to the $U(1)$ charge, is not exactly pertinent in this case.

Vice versa,  if the condensate   $\brc \psi\eta \ckt $ does occur,   and thus the system is in  Higgs phase,  then there is a  $U(1)$  symmetry which is spontaneously 
broken, and  a physical NG boson appears (see  Sec.~\ref{NG2}  above). This  NG boson {\it  can} appear as  an intermediate state of the two-point function of the associated currents,  in the leading $N$ planar graph.  

Indeed, for the large $N$ counting of Feynman diagrams in the $\psi\eta$  model, it is sufficient to remember \cite{Eichten:1985fs} that neither $\psi$ nor $\eta$ loops are suppressed by $\frac{1}{N}$: the standard $\frac{1}{N}$  suppression for the fundamental-fermion ($\eta$) loop inside a planar graph is compensated by a factor $\sim N$ from the flavor multiplicity.  
 This leads to the observation \cite{Eichten:1985fs}   that  the diagrams with  $B {\bar B}$,   $B {\bar B}B {\bar B}$, etc.,  intermediate states \footnote{$B$ are the hypothetical massless composite fermions  (``baryons") of  (\ref{baryons}),  $B \sim  \psi \eta \eta$. }
  are not suppressed  by  a  $\frac{1}{N}$ factor. 
  
\begin{figure}
\begin{center}
\includegraphics[width=4.8in]{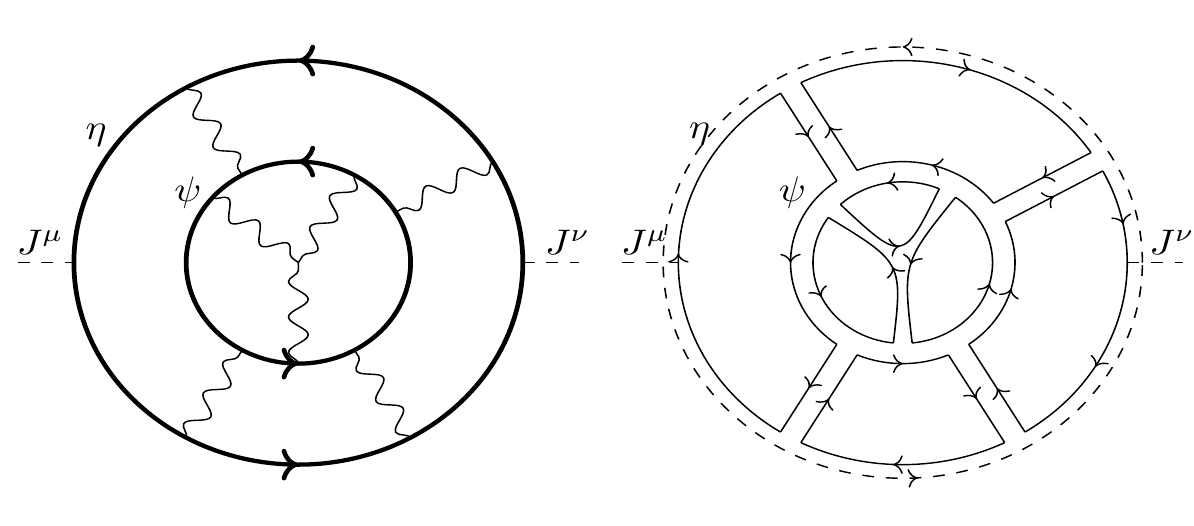}
\caption{\small  A leading $N$ planar  graph for the two-point  function 
$\brc 0 | T\{J_{\mu}(x)  J_{\nu}(0)  \}| 0 \ckt $   with  an  $\eta^* \psi^*  \psi\eta$ intermediate state (Fig.1a, left).  On the right figure (Fig. 1b) the color (full) and flavor (dashed) lines are
shown. The wavy lines inside the graph are the gluons.}
\label{Fig1}
\end{center}
\end{figure}

\begin{figure}[h!]
\begin{center}
\includegraphics[width=2.6in]{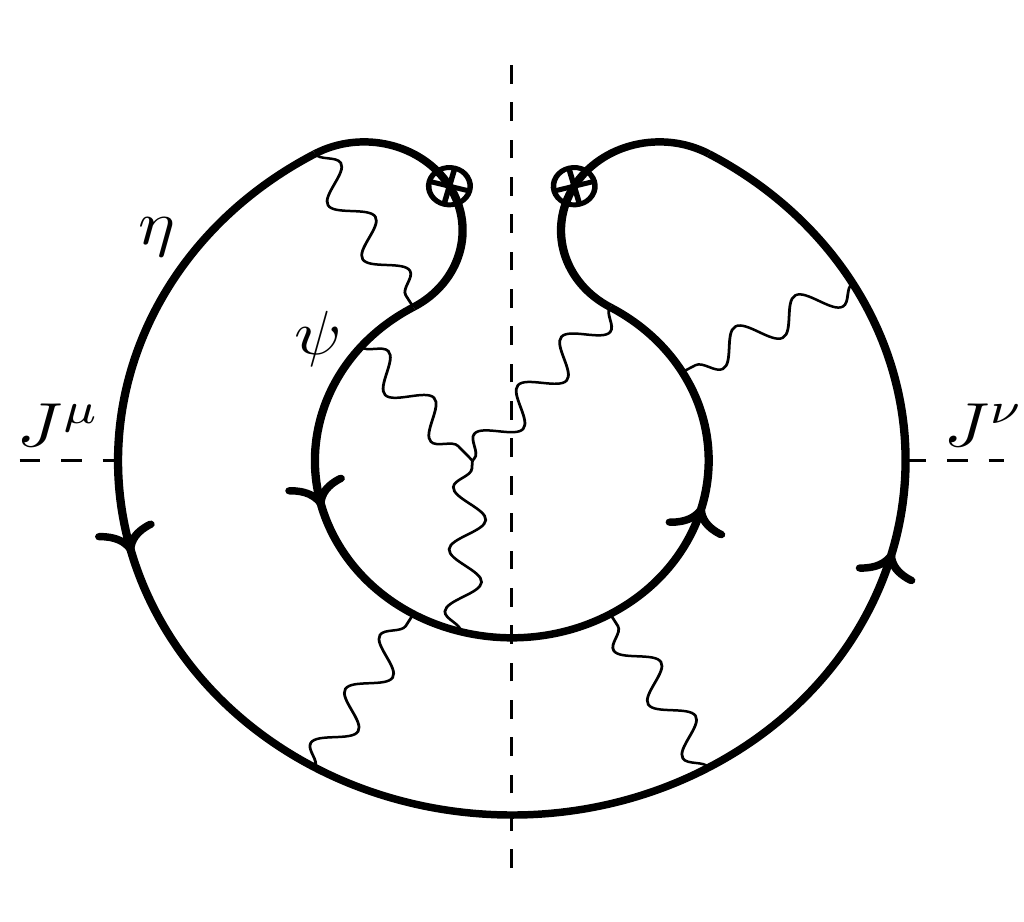}
\caption{\small The same graph as Fig. 1, with insertion of   $\brc \psi \eta \ckt$ condensates (shown with $\otimes$).  The intermediate state now corresponds to the  $U(1)$ NG boson  $\sim \psi \eta$ of 
Eq.~(\ref{pionpsieta}).    }
\label{Fig2}
\end{center}
\end{figure}

   For exactly the same reason,  neither are the  planar graphs with 
 the intermediate states   $\eta \psi  \psi^* \eta^*$,  $\eta \psi  \eta \psi   (\psi \eta)^* (\psi \eta)^*$,  and $\det U (\det U)^*$ \footnote{The $\psi$ and $\eta$ loops (rings)  in the graph must be ordered one inside the other, alternatively, in order to keep the leading $\frac{1}{N}$ order. Arbitrary gluon exchanges can be added   in a planar graph as in    Fig.~\ref{Fig1}. }.   A simplest such graph is shown  in  Fig.~\ref{Fig1}.  Allowing for  nonvanishing condensates  $\brc \psi\eta \ckt $  to occur in the diagram, one can see that   
 the  $U(1)$  NG boson does appear as an intermediate state in the  current two-point function, Fig.~\ref{Fig2}.  There is no difficulty  showing the same by using the gauge-invariant condensate and pion field, (\ref{gaugeinvCond}), even though it becomes more cumbersome to draw a picture.

 Recapitulating, a large N counting argument neither discriminates the possible phases  nor prohibits a 
  $U(1)$ symmetry to be broken spontaneously.      
 

\section{Multifield versus bi-fermion condensates}

One of the key elements of our argument based on the strong anomaly   is the statement  that the multifield condensate such as   the one  in the $\chi\eta$ model,  
\be      \brc    (\chi\eta)^{N-4} \chi\chi \ckt      \ne 0    \label{multicond}    \ee   
(see Eq.~(\ref{zeromodes}) for the exact meaning)    implies the  condensation of the component bi-fermion  scalars,  
\be   \brc \chi\eta \ckt\ne 0\;, \qquad     \brc \chi\chi \ckt\ne 0\;, \label{compocond}
\ee
where 
\be     \brc \chi\eta \ckt\ =  \brc   \chi_{[ij]} { \eta}^{B\, j}   \ckt  \sim    \, \Lambda^3 \delta_i^B \;, \qquad i, B=1,2,\ldots, N-4\;,  \label{compon1}
\ee
\be    \brc \chi\chi \ckt   =   \brc   \epsilon_{i_1 i_2 i_3 i_4} \chi^{i_1 i_2} \chi^{i_3 i_4} \ckt  \sim   \Lambda^3 \;, \qquad    N-3  \le  i_j  \le  N\;,   \label{compon2}
\ee
in an appropriate gauge and with flavor orientation.
A similar step is used in all other models considered here.  
As this point is central in our discussion, let us pose, before summarizing and concluding this paper, to make a few more comments.  
We will use the $\chi\eta$ model (the simplest of the GG models) for this purpose,  for concreteness and to keep the notations simple.

We have not proven that (\ref{compocond})  follows from   (\ref{multicond}).    We do believe this is correct, but our deduction is based on a collection of plausibility 
considerations and consistency checks:   dealing with a strongly-coupled gauge theories, to mathematically prove this type of derivation is not always an easy task.

Here are a few more  considerations, supporting our deduction. 

\begin{description}
 
\item[(a)]   Quark condensates in QCD versus $\chi\eta$ condensate:
 
In QCD, 
the quark condensate
\be   \brc {\bar \psi}_R  \psi_L \ckt \sim \Lambda^3 \ne 0
\ee
does not  imply 
\be \brc {\bar \psi}_R \ckt \ne 0\;, \qquad    \brc {\psi}_L \ckt \ne 0\;. \ee
However this is not   because  the
quarks are not gauge invariant,  but because they are fermions. 

But then   as we recalled in Sec. 3.1,   the multifield condensate
\be  \brc \det U \ckt =     \brc \det  {\bar \psi}_R  \psi_L    \ckt \ne 0
\ee
following from the strong-anomaly effective action, does seem to imply  the  bi-fermion scalar condensates,  $   \brc{\bar \psi}_R  \psi_L    \ckt  \ne 0$.
We  will further argue in Sec.~\ref{Summary} that  the scalar composites  ${\bar \psi}_R  \psi_L  $    in QCD  and    $\chi\eta$ in the GG model (and similarly $ \psi\eta$ in the BY model, etc.)
should be regarded as analogous objects.

We may  further add that if  the $SU(2)_{WS}\times U(1)_Y$   gauge interactions are taken into account,  the quark condensates   $   \brc{\bar \psi}_R  \psi_L    \ckt  \ne 0$
are no longer gauge-invariant objects \footnote{As is well known, the quark condensate breaks the Weinberg-Salam gauge symmetry dynamically,  even though only 
by a tiny amount, insufficient to  explain in itself the observed electroweak gauge symmetry breaking. }.   In this sense, too,     there is no qualitative  difference  between  the scalar composites  ${\bar \psi}_R  \psi_L  $    in QCD  and    $\chi\eta$ in the GG model.

\item[(b) ]  The standard-model Higgs doublet:

In the Weinberg-Salam model, the vacuum is at the gauge-invariant minimum of the potential, 
\be   \brc   \sum_{i=1}^2  \phi^{i \, *}  \phi^i  \ckt    \ne 0\;.   \label{WS1}
\ee
This is taken to mean a nonvanishing Higgs VEV, in an appropriate  gauge, 
\be      \brc \phi  \ckt =    \left(\begin{array}{c}v \\0\end{array}\right)\;, \qquad v\ne 0\;.     \label{WS2}
\ee
Our deduction   (\ref{multicond}) $ \leftrightarrow$ (\ref{compocond})    is analogous to this, even though here we have a dynamical, composite ``Higgs" scalars.   
One can  argue against the use of this analogy, on the basis  that the Weinberg-Salam model is a weakly coupled theory, while  here we have a strongly coupled one. We believe  that this difference 
(the former described by a potential, the latter not) is not essential.   What excludes the conclusion that the Weinberg-Salam model with  (\ref{WS1}) is in confinement phase with 
\be   \brc   \sum_1^2  \phi^{i\, *}  \phi^i  \ckt    \ne 0\;,   \qquad   \brc \phi    \ckt =0\;,\label{WS3} 
\ee
is  the mass spectrum   ($W$, $Z$, $\gamma$, etc).  The attempt to  rewrite the whole Weinberg-Salam  model as a  confining theory \cite{Abbott}
fails to reproduce  exactly  the  mass spectra   (though it does almost)   of the standard Higgs phase description.  

Our observation here   is similar:  even though the global symmetry may look  the same  in two ``complementary" descriptions  of the  $\chi\eta$ model   \footnote{Remember that the ``complementarity"  aspect is specific to the $\chi\eta$ model,   not shared by any other  GG or BY models.}, the exact mass spectra are probably  different, as we pointed out in Sec.~\ref{complementarity}.   

%
%
%
%

\item[(c)] Consistency: 

A more indirect, consistency  argument is the following.  Let us assume the multifield condensate  (\ref{multicond})  forms  but with
\be  \brc \chi\eta\ckt=    \brc \chi\chi\ckt=   0\;,    \label{incons}
\ee
and with
the full (nonanomalous) symmetry of the $\chi\eta$   system,
\be 
 SU(N)_{\rm c} \times  SU(N- 4) \times U(1)_{\chi\eta}\;
\ee
intact.  The low-energy system is then described, as the conventional 't Hooft anomaly argument suggests,  by the set of massless  baryons  
 (\ref{massless}).  But as there is no way to describe  the mutifermion condensate   (\ref{multicond})   in terms of these massless fields,   we conclude that the assumption (\ref{incons})    is inconsistent.

One may ask why the confining phase with baryons cannot have some 
  composite scalar  $\sim  (\chi\eta)^{N-4} \chi\chi$   entering strong anomaly effective action.   To answer this, we recall 
   our criterion:  
 {\it    the infrared degrees of freedom of the assumed phase must be able to  describe the strong anomaly effectively at low energies.  }  
  Just as  the NG boson fields  do  in the Higgs phase  of the GG models, and as they do in the confining phase in QCD.  In the BY models,  the massless baryons {\it  and}  NG bosons together 
  saturate the argument of the logarithm, in the dynamical Higgs phase, as we discussed in Sec.~\ref{sec:stronganomaly}. 
  
If one adopts this criterion,  then  the confining, symmetric phase is  disfavored  (inconsistent)   in {\it   all}  of the GG and BY models.

\end{description}

\section{Summary and Discussion
 \label{Summary}}

Consideration of the strong-anomalies, not always taken into account in the study of dynamics of chiral gauge theories, turns out to provides us with a powerful new argument supporting  the conclusion of \cite{BKL2,BKL4}:  the system is in dynamical Higgs phase    in the Bars-Yankielowicz and Georgi-Glashow models. 
The essential requirement we make is that it should be possible to write a strong-anomaly effective action by
utilizing only the massless degrees of freedom, assumed to be present in the infrared. This simple requirement is naturally satisfied in all models considered, if assumed to be in the 
dynamical Higgs vacuum with bifermion condensates.  On the contrary,  in the putative, confinement vacua with massless composite fermions (baryons) as the only  infrared degrees of freedom, we found no way of writing the correct strong-anomaly effective action.  In other words,  it seems that  in these latter vacua, the anomalous $U(1)$ symmetry(ies)  cannot be properly realized in the infrared.  

The fact that the matter fermions in our models have larger multiplicities than the quarks  in QCD  (for instance  $\eta$  in the $\psi\eta$ model has the flavor multiplicity, $N+4 \sim N$  at large $N$), may mean that  the mass of the  anomalous $U(1)$ NG bosons will turn out to be  of the oder of $\Lambda$. Even though we do write the strong-anomaly effective action in terms of the massless degrees of freedom of the presumed  infrared theory,  the fact that its solution (the minimum of the effective potential) is at $\mu \sim \Lambda$, means that the answer cannot be regarded as reliable quantitatively.     We are however interested here  only  in a {\it qualitative} question 
of the validity of  confinement vs dynamical Higgs phase in the infrared,  and  we do believe that  our discussion of Sec.~\ref{sec:stronganomaly}  makes reasonable sense.

The fact that   both  mixed anomalies  \cite{BKL2,BKL4}  and  the strong-anomaly effective action (this paper)   imply dynamical Higgs phase in chiral BY and GG  models 
 -  is actually not a pure coincidence.  Both indeed arise by taking properly the strong chiral $U(1)$ anomalies into account.  

Certain analogies and contrasts between the strong-interaction dynamics of vectorlike and chiral gauge theories seem to emerge  from these discussions.  
As  the  ``confining, flavor symmetric"  vacua  have been shown to be disfavored,  we  assume below that  these chiral gauge theories
are indeed in dynamical Higgs phase.

And let us compare  the physics of the 
$\psi\eta$, $\chi\eta$  and of more general   BY and GG models,  with that of 
 the standard QCD  with $N_{\rm f}$  light flavors of  quarks and antiquarks.    Both in QCD and in these chiral gauge theories,   the theory without the matter fermions is the same,  pure $SU(N)$ Yang-Mills theory.  We have here nothing new to add to the known and unknown
properties  of the pure $SU(N)$ dynamics.
Our interest  here is the role the light matter fermions play in determining  the infrared dynamics, and how the resulting phase(s)  depend on the types of the matter fermions present. 
 
   In many senses,  the bifermion condensates such as  $U=  \psi\eta$  in the $\psi\eta$ model  (and $U= \chi\eta$ and  $\chi\chi$ condensates in the $\chi\eta$  model), are a good analogue of the quark condensate
   $U= {\bar \psi}_R \psi_L  $ in QCD.   This ``analogy" is based on the fact that  all of these composite scalar fields enter the strong-anomaly effective action in the same way, as
   \be    {\hat L}  =  \frac{i}{2}    q(x)  \log   \det   U/ U^{\dagger}\;, \qquad      q(x)  =  \frac{g^2}{32\pi^2}  F_{\mu\nu}^a  {\tilde F}^{a, \mu\nu}\;.
\label{anompsietaAll}
\ee 
(see Sec.~\ref{sec:stronganomaly} for more careful discussions).
And in all cases  this implies condensation, $\brc U \ckt \propto {\mathbf  1}$,   although  it's meaning depends on the system:  the color-flavor locked Higgs phase in the chiral models, and  the chiral-symmetry-breaking confining vacuum  in QCD.

Also,   in a more general Bars-Yankielowicz  models,  there are two natural bifermion condensation channels   ($ (\cdot)$ meaning an $SU(N)$  singlet):
 \bea   &   \psi \Big(\raisebox{-2pt}{\yng(2)}\Big) \,  \eta \Big(\bar{\raisebox{-2pt}{\yng(1)}}\Big)      \qquad  \  & {\rm forming}    \qquad      \raisebox{-2pt}{\yng(1)}\;,  \nonumber    \\
   &   \qquad   \xi \Big({\raisebox{-2pt}{\yng(1)}}\Big) \, \eta \Big(\bar{\raisebox{-2pt}{\yng(1)}}\Big)    \qquad   \ & {\rm forming}  \qquad  (\cdot)  \;:
 \eea
 the gluon-exchange strengths in the two channels  
are, respectively, proportional to \\
   $  -      \frac{ (N+2)(N-1)}{N} $  and  
    $ -       \frac{N^2-1}{N} $\;.
     The  $\psi \eta$ channel  is slightly more attractive,  but the strengths are  identical in the large $N$ limit. Note that in our notation  $\brc  \xi \eta  \ckt$ has  the same quantum numbers as  $\brc {\bar \psi}_R \psi_L\ckt $ in QCD.
     Similarly for the comparison between the  condensates   $\brc \chi \eta \ckt$ and   $\brc  \xi \eta  \ckt$ in the general Georgi-Glashow models.
These considerations,  which are based on rather na\"ive  MAC \cite{Raby}  like  idea and therefore are not very rigorous,  nevertheless show that  the  quark condensates in QCD   and  the bifermions condensates in the chiral gauge theories under study,  are really on a very ``similar"  footing.  

Of course,  the fact that the quark condensate $\brc {\bar \psi}_R \psi_L\ckt $  is a color singlet,  $SU(N_{\rm f})_L\times SU(N_{\rm f})_R$ flavor matrix,     whereas  $\brc \psi\eta \ckt$  is  in a  color-flavor bifundamental form, breaking the color completely,  makes a world of differences in other aspects of these (vectorlike or chiral)  systems.   Most importantly, the existence of colored NG bosons in the $\psi\eta$  (or in the $\chi\eta$)  model,  see Sec.~\ref{colored},  means that these are coupled linearly to the color gauge bosons, and via Englert-Brout-Higgs mechanism make them massive.   These processes are absent in QCD, all NG bosons being color singlets.  It is in this sense  that we talk about confinement phase in QCD,
in spite of the fact that the linear potential between any two test particles can be  always flattened  by the pair production of quarks from the vacuum,  
  and Higgs phase in the chiral models. 

Also, the mass spectra get arranged quite differently in QCD and in the chiral models discussed here.  Apart from the pattern of certain degenerate massive vector bosons
(see Sec.~\ref{complementarity}),  the {\it  massless spectrum} exhibits striking differences.   In all chiral gauge theories considered here,   it 
 contains in general {\it both} a number of composite fermions (baryons) as well as some composite scalars (pions),
  a feature not shared by massless QCD.   In other words, the way the chiral symmetries of the UV theory are realized in the IR, is distinctly different.
  
Perhaps a closer analogy -  from a formal point of view - between the vectorlike theories and chiral theories,   comes from the consideration of 
color superconductivity in the high-density limit of QCD \cite{Alford1997,Alford1998}. In such a situation the dynamics of QCD is believed to be such that the colored di-quark condensates
form,
    \be     \brc \psi_L  \psi_L  \ckt   \ne 0\;,   \qquad     \brc \psi_R  \psi_R  \ckt \ne 0\;.
    \ee
   In particular, for $N_{\rm f}=3$, they are condensates of color-flavor diagonal forms,  somewhat similar to $\brc \psi\eta \ckt$  or  $\brc \chi\eta \ckt$ 
in the chiral theories discussed here, although the details of the dynamics can obviously be quite different. 

What lessons should one draw from these discussions?  Clearly, there are both  similarities and differences 
 between the dynamics of QCD and that of the chiral gauge theories  discussed here.   The consideration of the strong anomaly effective actions discussed  here  seems   to lead us to clearer meaning of these comparisons, and with that,  to point to a better understanding of the dynamics of 
  strongly-coupled chiral gauge theories.

\section*{Acknowledgments}  

We thank Gabriele Veneziano for bringing \cite{Vene}  to our attention and for a comment, and David Tong for exchanges of views on some of the topics discussed here.
  The work is supported by the INFN special 
research initiative grant, ``GAST" (Gauge and String Theories).

\appendix

   \section{Chirally symmetric phase   in the $\psi\eta$ model
\label{conf1} } 
    
An interesting possibility  for the $\psi\eta$ model is  that no condensates form, the system confines and the flavor symmetry is unbroken \cite{BY}.
    The candidate massless degrees of freedom in the IR  are  $\frac{(N+4)(N+3)}{2}$ ``baryons", 
      \be     B^{[AB]}=    \psi^{\{ij\}}  \eta_i^A  \eta_j^B \;,\qquad  A,B=1,2, \ldots, N+4\;,     \label{baryons}
\ee
antisymmetric in  $A \leftrightarrow B$.
All the $SU(N+4)\times U(1)$ anomalies are saturated by  $ B^{[AB]}$ as can be seen by using the data in Table~\ref{Simplest0}. 
The discrete anomaly   $(\mathbbm Z_{N+2})_{\psi} -  [SU(N)]^2$  is also matched as can be easily checked,  and all other discrete anomalies are also matched as a consequence.
\begin{table}[h!t]
  \centering 
  \begin{tabular}{|c|c|c |c|c|  }
\hline
$ \phantom{{{   {  {\yng(1)}}}}}\!  \! \! \! \! \!\!\!$   & fields  &  $SU(N)_{\rm c}  $    &  $ SU(N+4)$     &   $ { U(1)_{\psi\eta}}   $  \\
 \hline 
  \phantom{\huge i}$ \! \!\!\!\!$  {\rm UV}&  $\psi$   &   $ { \yng(2)} $  &    $  \frac{N(N+1)}{2} \cdot (\cdot) $    & $   N+4$    \\
 & $ \eta^{A}$      &   $  (N+4)  \cdot   {\bar  {\yng(1)}}   $     & $N \, \cdot  \, {\yng(1)}  $     &   $  - (N+2) $ \\
   \hline     
 $ \phantom{ {\bar{   { {\yng(1,1)}}}}}\!  \! \! \! \! \!\!\!$  {\rm IR}&    $ B^{[AB]}$      &  $  \frac{(N+4)(N+3)}{2} \cdot ( \cdot )    $         &  $ {\yng(1,1)}$        &    $ -N    $   \\
\hline
\end{tabular}
  \caption{\footnotesize  Chirally symmetric ``confining"  phase of the  $\psi \eta $  model.   As in other Tables of the text, 
the multiplicity, charges and the representation are shown for each  set of fermions. $(\cdot)$ stands for a singlet representation.
}\label{Simplest0}
\end{table}

\section{Color-flavor locked Higgs phase  in the $\psi\eta$ model  \label{Higgs1}  }

Another possibility, in the $\psi\eta$ model,  is  that of  a color-flavor locked phase \cite{ADS,BKS}, with 
\be    \brc  \psi^{\{ij\}}   \eta_i^B \ckt =\,   C \,  \Lambda^3   \delta^{j B}\;,   \qquad   j, B=1,2,\dots  N\;,    \label{cflocking}
\ee
in which the symmetry is  reduced to 
\beq
SU(N)_{\rm cf} \times  SU(4)_{\rm f}  \times U(1)^{\prime} \,.   \label{isnot} 
    \eeq
    A subset of the same baryons   ($B^{[A_1 B_1]}$ and   $B^{[A_1 B_2]}$ in the notation of Table~\ref{SimplestBis})
    saturate all of the triangles associated with the reduced symmetry group, see Table~\ref{SimplestBis}.  
The  massless  degrees of freedom are  $\tfrac{N^2+7N}{2}$ massless baryons  $B^{[A_1 B]}$   and   $8N+1$  NG bosons.
   To reproduce correctly the strong anomaly in the IR, however,  another condensate $  \brc { B}  { B}   \ckt \sim  \brc  \psi \eta \eta \psi \eta \eta    \ckt $ 
 and another set of  massless baryon  $B^{[A_2 B_2]}$ (see Table~\ref{SimplestBis}) are needed.
This does not alter  neither the symmetry breaking pattern (\ref{isnot}), nor the anomaly matching. See Sec.~\ref{strongpsieta} for more discussion.
 \begin{table}[h!t]
  \centering 
  \begin{tabular}{|c|c|c |c|c|  }
\hline
$ \phantom{{{   {  {\yng(1)}}}}}\!  \! \! \! \! \!\!\!$   & fields   &  $SU(N)_{\rm cf}  $    &  $ SU(4)_{\rm f}$     &   $  U^{\prime}(1)   $  \\
 \hline
   \phantom{\huge i}$ \! \!\!\!\!$  {\rm UV}&  $\psi$   &   $ { \yng(2)} $  &    $  \frac{N(N+1)}{2} \cdot   (\cdot) $    & $   1  $    \\
 & $ \eta^{A_1}$      &   $  {\bar  {\yng(2)}} \oplus {\bar  {\yng(1,1)}}  $     & $N^2 \, \cdot  \, (\cdot )  $     &   $ - 1 $ \\
&  $ \eta^{A_2}$      &   $ 4  \cdot   {\bar  {\yng(1)}}   $     & $N \, \cdot  \, {\yng(1)}  $     &   $ - \frac{1}{2}  $ \\
   \hline 
   $ \phantom{{\bar{ \bar  {\bar  {\yng(1,1)}}}}}\!  \! \!\! \! \!  \!\!\!$  {\rm IR}&      $ B^{[A_1  B_1]}$      &  $ {\bar  {\yng(1,1)}}   $         &  $  \frac{N(N-1)}{2} \cdot  (\cdot) $        &    $   -1 $   \\
       &   $B^{[A_1 B_2]}$      &  $   4 \cdot {\bar  {\yng(1)}}   $         &  $N \, \cdot  \, {\yng(1)}  $        &    $ - \frac{1}{2}$   \\
       &   $B^{[A_2 B_2]}$      &  $   6  \cdot (\cdot )    $         &  $ {\yng(1,1)}  $        &    $0$   \\
\hline
\end{tabular}  
  \caption{\footnotesize   Color-flavor locked phase in the $\psi \eta$ model.
  $A_1$ or $B_1$  stand for the first $N$ flavors ($A_1,B_1=1,2,\ldots, N$),  whereas  $A_2$ or $B_2$ run  over 
   the rest of the flavor indices, $N+1,\ldots, N+4$.  The set of potentially massless baryons  $B^{[A_2 B_2]}$,  which were not explicitly taken into account in \cite{ADS,BKL4},  do not contribute to 
   $SU(N)_{\rm cf} \times  SU(4)_{\rm f} \times U^{\prime}(1)$ anomalies. 
   }\label{SimplestBis}
\end{table}

\section{Chirally symmetric phase   in the $\chi\eta$ model   \label{conf2}}

    Let us first examine the possibility that no condensates form, the system confines and the flavor symmetry is unbroken
   \cite{Dimopoulos:1980hn}.
 The massless baryons are   
 \be      B^{\{CD\}} = \chi_{[ij]} \,  {\eta}^{i\, C}    {\eta}^{j\, D}   \;, \qquad   C,D=1,2,\ldots (N-4)\;,  \label{massless}\ee
 symmetric in $C \leftrightarrow D$.   
 The matching of the anomalies can be read off Table~\ref{01model}.
\begin{table}[h!t]
  \centering 
  \begin{tabular}{|c|c|c |c|c|  }
\hline
 $ \phantom{{{   {  {\yng(1)}}}}}\!  \! \! \! \! \!\!\!$   & fields   &  $SU(N)_{\rm c}  $    &  $ SU(N-4)$     &   $ U(1)_{\chi\eta}   $  \\
 \hline 
  $ \phantom{{\bar{ \bar  {\bar  {\yng(1,1)}}}}}\!  \! \!\! \! \!  \!\!\!$  {\rm UV}&  $\chi$   &   ${\bar  { \yng(1,1)}}   $  &    $  \frac{N(N-1)}{2} \cdot (\cdot) $    & $N-4$    \\
& $ {\eta}^{A}$      &   $  (N-4)  \cdot   { {\yng(1)}}   $     & $N \, \cdot  \, {\yng (1)}  $     &   $ - (N-2)  $ \\
   \hline     
  \phantom{\huge i}$ \! \!\!\!\!$  {\rm IR}&    $ B^{\{AB\}}$      &  $  \frac{(N-4)(N-3)}{2} \cdot ( \cdot )    $         &  $ {\yng(2)}$        &    $ - N $   \\
\hline
\end{tabular}
  \caption{\footnotesize  Confinement and unbroken symmetry in the $\chi\eta$ model} 
  \label{01model}
\end{table}


\section{Color-flavor locked Higgs vacuum  in the $\chi\eta$ model  \label{Higgs2} }

It was pointed out \cite{ADS} that this system may instead develop a condensate of the form
\be   \brc   \chi_{[ij]} { \eta}^{B\, j}   \ckt  = C   \, \Lambda^3 \delta_i^B \;, \qquad i, B=1,2,\ldots, N-4\;,  \label{cfl01}
\ee
namely,
\be   {\bar  {\yng(1,1)} }  \otimes  \yng (1)  \to    {\bar  {\yng(1)} }  \oplus  \ldots \;.
\ee
The symmetry is broken to 
\beq
   SU(N-4)_{\rm cf}  \times     SU(4)_{\rm c} \times U(1)'  \,.
    \label{b270}
\eeq
The massless baryons (\ref{massless})  saturate all the anomalies associated with $SU(N-4)_{\rm cf}  \times  U(1)^{\prime}$.
There remains the  $\chi_{i_2 j_2}$  fermions which remain massless and strongly coupled to the $SU(4)_{\rm c}$.  We may assume that 
$SU(4)_{\rm c}$ confines, and the condensate
\be  \brc  \chi \chi \ckt \ne 0\;, 
\ee
in
 \be    {\bar {  \yng(1,1) }} \otimes  {\bar {  \yng(1,1) }}  \, \to \,  {\bar { \yng(1,1,1,1) }}  \, \oplus  \ldots \;,
\ee
forms and $\chi_{i_2 j_2}$ acquire dynamically mass. 
Assume that the massless baryons are:
\be      B^{\{A B\}} = \chi_{[ij]} \,  {\eta}^{i\, A}   {\eta}^{j\, B}   \;, \qquad   A,B=1,2,\ldots (N-4)\;, 
\ee
 the saturation of all of the triangles associated can be seen in Table~\ref{SimplestAgain2}.  
\begin{table}[h!t]
  \centering 
  \begin{tabular}{|c|c|c |c|c|  }
\hline
$ \phantom{{{   {  {\yng(1)}}}}}\!  \! \! \! \! \!\!\!$   & fields     &  $ SU(N-4)_{\rm cf} $     &   $ U(1)^{\prime} $     &  $SU(4)_{\rm c}  $     \\
 \hline
   $ \phantom{{\bar{ \bar  {\bar  {\yng(1,1)}}}}}\!  \! \!\! \! \!  \!\!\!$  {\rm UV}&  $\chi_{i_1 j_1}$     &    $  {\bar  { \yng(1,1)}}   $    & $N$   &   $\frac{(N-4)(N-5)}{2}\cdot (\cdot)  $ \\
 &  $\chi_{i_1 j_2}$   &    $  4   \cdot {\bar  { \yng(1)}} $    & $\frac{N}{2}$   &   $ (N-4) \cdot {\bar  { \yng(1)}}   $     \\
 &$\chi_{i_2 j_2}$   &    $  \frac{4 \cdot 3}{2} \cdot (\cdot) $    & $0$    &   ${\bar  { \yng(1,1)}}   $     \\
& $ {\eta}^{i_1,A}$          & $\yng(2) \oplus \yng(1,1)$     &   $ - N $    &   $  (N-4)^2  \cdot  (\cdot)   $  \\
 & $ {\eta}^{i_2,A}$         & $4\, \cdot  \, {\yng (1)}  $     &   $ - \frac{N}{2} $     &   $  (N-4)  \cdot  \yng(1)  $  \\
   \hline 
     \phantom{\huge i}$ \! \!\!\!\!$  {\rm IR}&     $ B^{\{AB\}}$        &  $ {\yng(2)}$        &    $ - N $     &  $  \frac{(N-4)(N-3)}{2} \cdot ( \cdot )    $      \\
\hline
\end{tabular}
  \caption{\footnotesize  Color-flavor locking  in the $\chi\eta$ model.    The color index $i_1$ or $j_1$  runs up to $N-4$ and the rest is indicated by $i_2$ or $j_2$.}\label{SimplestAgain2}
\end{table}
 The complementarity \cite{Fradkin} apparently works here, in the sense that the massless sector of the dynamical  Higgs phase has the same 
 $SU(N-4)\times U(1)$ symmetry.  See Sec.~\ref{complementarity} for more discussion. 
%
%

\section{Confining phase with unbroken global symmetries  of the  BY  models     \label{conf3}}

The matter fields of the BY model are:
\begin{table}[h!t]
  \centering 
  \small{\begin{tabular}{|c|c|c|c|c|c|  }
\hline
\su      &  $SU(N)_{\rm c}  $    &  $ SU(N+4+p)$    &  $ SU(p)$     &   $ {U}(1)_{\psi\eta}   $  &   $ {U}(1)_{\psi\xi}   $  \\
\hline 
\sbu  $\psi$   &   $ { \yng(2)} $  &    $  \frac{N(N+1)}{2} \cdot (\cdot) $    & $   \frac{N(N+1)}{2} \cdot (\cdot)  $  & $N +4 +p$ & $p $  \\
  $ \eta$      &   $  (N+4+p)  \cdot   {\bar  {\yng(1)}}   $     & $N  \cdot  {\yng(1)}  $     & $N (N+4+p) \, \cdot  (\cdot)   $   &$-(N+2)$&$0$\\ 
$ \xi$      &   $  p \cdot   {  {\yng(1)}}   $     & $N p \, \cdot  (\cdot)   $     &  $ N  \cdot   { {\yng(1)}} $ &$0$&$-(N+2)$ \\
\hline   
\end{tabular}}
  \caption{\footnotesize The multiplicity, charges and the representation are shown for each  set of fermions in the BY model.  
}\label{suv}
\end{table}

%
%

   The candidate massless composite fermions  for the Bars-Yankielowitz  models are the left-handed gauge-invariant fields:
      \be    
 {({ B}_{1})}^{[AB]}=    \psi^{ij}   \eta_i^{A}  \eta_j^{B}\;,
\qquad {({ B}_{2})}^{a}_{A}=    \bar{\psi}_{ij}  \bar{\eta}^{i}_{A}  \xi^{j,a} \;,
\qquad {({ B}_{3})}_{\{ab\}}=    \psi^{ij}  \bxi_{i,a}  \bxi_{j,b}  \;,
\label{baryons100}
\ee
the first is anti-symmetric in $A \leftrightarrow B$ and the third is  symmetric in $a \leftrightarrow b$; their charges are given in Table  \ref{sir}.
Writing explicitly also the spin indices they are 
 \bea    
& {({ B}_{1})}^{[AB], \alpha}=   \frac{1}{2} \epsilon_{\beta \gamma} \psi^{ij, \beta}  \eta_i^{A, \gamma}  \eta_j^{B, \alpha}  +  \frac{1}{2} \epsilon_{\beta \gamma} \psi^{ij, \beta}  \eta_i^{A, \alpha}  \eta_j^{B, \gamma}   
\;, &\nn \\
& {({ B}_{2})}^{a, \alpha}_{A}=     \epsilon_{\dot{\alpha}\dot{\beta} }  \bar{\psi}_{ij}^{\dot{\alpha}}  \bar{\eta}^{i,\dot{\beta}}_{A}  \xi^{j,a,\alpha} \;,
\qquad {({ B}_{3})}_{\{ab\}}^{\alpha}=    \epsilon_{\dot{\beta} \dot{\gamma}} \psi^{ij, \alpha}  \bxi_{i,a}^{\dot{\beta}}  \bxi_{j,b}^{\dot{\gamma}} 
\;: &
\label{b}
\eea
all  transforming under the  $\{\tfrac{1}{2}, 0\}$ representation of the Lorentz group.
\begin{table}[h!t]
  \centering 
  \small{\begin{tabular}{|c|c|c |c|c| c|c| }
\hline
\su      &  $SU(N)_{\rm c}  $    &  $ SU(N+4+p)$    &  $ SU(p)$     &   $ {U}(1)_{\psi\eta}   $  &   $ {U}(1)_{\psi\xi}   $  \\
  \hline     
 \sbuu     $ {{ B}_{1}}$      &  $   \frac{(N+4+p)(N+3+p)}{2} \cdot ( \cdot )    $         &  $ {\yng(1,1)}$        &    $  \frac{(N+4+p)(N+3+p)}{2} \cdot ( \cdot )     $   &  $-N+p $ & $p$ \\
  \hline     
  \sbbu   $ {{ B}_{2}}$   &    $ (N+4+p) p\cdot ( \cdot )$     &       $  p\cdot   \bar{\yng(1)}$   &     $ (N+4+p)  \cdot {\yng(1)}$      & $-(p+2)$ & $-(N+p+2)$ \\
  \hline     
   \sbbu  ${{ B}_{3}}$     &  $ \frac{p (p+1)}{2}  \cdot ( \cdot )   $    &    $ \frac{p (p+1)}{2}  \cdot ( \cdot )   $       &    $ \bar{\yng(2)}$       & $N+4+p$ & $2N +4 + p$\\
\hline
\end{tabular}}
  \caption{\footnotesize  Chirally symmetric phase of the  BY  model. 
}\label{sir}
\end{table}
Table \ref{suvsir} summarizes the anomaly matching checks \cite{BKL4} 
 via comparison between Table~\ref{suv} and Table~\ref{sir}.
\begin{table}[h!t]
\centering
 \tiny{\begin{tabular}{|c|c|c|}
\hline
\su      &  UV    &  IR  \\
\hline 
\su $ SU(N+4+p)^3$     &   $ N $     &  $ N+p -p $ \\
\su $ SU(p)^3$     &   $ N $     &  $ N+4+p -(p+4) $ \\
\su $ SU(N+4+p)^2 - {U}(1)_{\psi\eta}$     &   $ - N(N+2)  $     &  $ -(N+2+p)(N-p) -p(p+2) $ \\
\su $ SU(N+4+p)^2- {U}(1)_{\psi\xi}$     &    $ 0 $     &  $  (N+2+p)p -p(N+p+2)  $ \\
\su $ SU(p)^2- {U}(1)_{\psi\eta}$     &   $ 0 $     &  $ -(N+4+p)(p+2) +(p+2)(N+p+4)  $ \\
\su $ SU(p)^2- {U}(1)_{\psi\xi}$     &   $ - N(N+2)   $     &  $   -(N+4+p)(N+p+2) +(p+2)(2N+p+4)$ \\
\su $ {U}(1)_{\psi\eta}^3$     &   $ \frac{N(N+1)}{2}(N+4+p)^3 - N(N+4+p) (N+2)^3 $     &  $ - \frac{(N+4+p)(N+3+p)}{2} (N-p)^3 -(N+4+p) p (p+2)^3 + $ \\\
\su      &    
&  $ +  \frac{p (p+1)}{2}(N+4+p)^3$ \\
\su $ {U}(1)_{\psi\xi}^3$     &   $ \frac{N(N+1)}{2}p^3 - N p (N+2)^3  $     &  $    \frac{(N+4+p)(N+3+p)}{2} p^3 -(N+4+p) p (N+p+2)^3  +$ \\
\su    &     
 &  $    \  +  \frac{p (p+1)}{2}(2N+4+p)^3$ \\
\su $ {\rm Grav}^2-{U}(1)_{\psi\eta} $     &  $ \frac{N(N+1)}{2}(N+4+p)  - N(N+4+p) (N+2)  $     &  $ - \frac{(N+4+p)(N+3+p)}{2} (N-p)  -(N+4+p) p (p+2) + $ \\
\su      &   
&  $ +  \frac{p (p+1)}{2}(N+4+p) $ \\
\su $ {\rm Grav}^2-{U}(1)_{\psi\xi}$       &   $\frac{N(N+1)}{2}p - N p (N+2)  $     &  $    \frac{(N+4+p)(N+3+p)}{2} p -(N+4+p) p (N+p+2)  +$ \\
\su    &     
&  $    \  +  \frac{p (p+1)}{2}(2N+4+p)$ \\
\su $ SU(N+4+p)^2-({\mathbbm Z}_{N+2})_{\psi}  $     &  $0$    & $N+2+p-p = 0 \ {\rm mod} \ N+2$    \\
\su $ SU(p)^2- ({\mathbbm Z}_{N+2})_{\psi} $     &   $0$    & $-(N+4+p) + p+2  = 0 \ {\rm mod} \ N+2 $     \\
\su $ {\rm Grav}^2-({\mathbbm Z}_{N+2})_{\psi}  $     &  $1$      & $1-1+1$  \\
\hline   
\end{tabular}}
  \caption{\footnotesize  Anomaly matching checks for the IR chiral symmetric phase of the BY  model.    For $N$ odd, the last three equalities  
  are consequences of other equations.
}\label{suvsir}
\end{table}

\section{Dynamical Higgs phase in the BY  models   \label{Higgs3}}

The broken phase for the  simplest of this class, the $\psi \eta$ model,  has also been extensively in the main text,  
see also  \cite{ADS,BKS, BKL2, BKL4}. 
Something interesting happens for  models with $p > 0$ additional pairs of 
fermions in the fundamentals   $(\eta, \xi)$. Now there is another channel, $  \xi \eta $, which is gauge invariant and charged under the flavor group. We thus have a competition between two possible symmetry breaking channels, $\psi \eta$ and $  \xi \eta $. 
We assume that both  condensates occur in the following way:
\bea
&&  \brc  \psi^{ij}   \eta_i^B \ckt =\,   c_{\psi\eta} \,  \Lambda^3   \delta^{j B}\ne 0\;,   \qquad   j,B=1,\dots,  N\;,     \nonumber \\
&& \brc  \xi^{i,a}   \eta_i^A \ckt =\,   c_{\eta\xi} \,  \Lambda^3   \delta^{aA}\ne 0\;,   \qquad  a = 1,\dots, N\;,  \quad  A=N+1,\dots, N+ p \;, 
  \eea
where $\Lambda$ is the renormailization-invariant scale dynamically generated by the gauge interactions and $ c_{\eta\xi} ,  c_{\psi\eta} $ are coefficients both of order one. 
According to the tumbling scenario \cite{Raby}, the first condensate to occur is in  the maximally attractive channel (MAC).  The strengths of the one-gluon exchange potential for the two channels
 \bea   &   \psi \Big(\raisebox{-2pt}{\yng(2)}\Big) \,  \eta \Big(\bar{\raisebox{-2pt}{\yng(1)}}\Big)      \qquad  \  & {\rm forming}    \qquad      \raisebox{-2pt}{\yng(1)}\;,  \nonumber    \\
&  \quad  \   \xi \Big({\raisebox{-2pt}{\yng(1)}}\Big) \, \eta \Big(\bar{\raisebox{-2pt}{\yng(1)}}\Big)    \qquad   \ & {\rm forming}  \qquad  (\cdot)  \;, 
 \eea
are, respectively,
 \bea      \frac{N^2-1}{2N}  -      \frac{ (N+2)(N-1)}{N}  -   \frac{N^2-1}{2N}  &=&  -      \frac{ (N+2)(N-1)}{N} \;,\nonumber \\
    0 -   2      \frac{N^2-1}{2N}    &=&  -       \frac{N^2-1}{N}  \;.
 \eea
So the  $ \psi \eta$ channel  is slightly more attractive, but such a perturbative argument is not really significant and we assume here that both types of condensates are formed.     

The resulting pattern of symmetry breaking is
\bea
&& SU(N)_{\rmc}  \times   SU(N+4+p)_{\eta}  \times  SU(p)_{\xi}  \times  U(1)_{\psi\eta}\times  U(1)_{\psi\xi} \nn \\
&&  \xrightarrow{\brc  \xi   \eta \ckt , \brc  \psi \eta \ckt}      SU(N)_{{\rm cf}_{\eta}}  \times   SU(4)_{\eta}  \times  SU(p)_{\eta\xi}  \times  U(1)_{\psi \eta}^{\prime} \times  U(1)_{\psi \xi}^{\prime} \;.
\label{symbresBis}
\eea
At the end the color gauge symmetry is completely (dynamically) broken, leaving  color-flavor diagonal  $SU(N)_{{\rm cf}_{\eta}} $ symmetry.
$U(1)_{\psi \eta}^{\prime}$ and $U(1)_{\psi \xi}^{\prime}$ are  combinations respectively  of $U(1)_{\psi \eta}$ and $U(1)_{\xi \eta}$ with  the element of $ SU(N+4+p)_{\eta}$ generated by 
\be   t_{SU(N+4+p)_{\eta}}= \left(\begin{array}{c|c|c}
(-\alpha (p+2) - p\beta) {\bf 1}_{N\times N}&&\\
\hline
&\frac{\alpha(N-p) - \beta p}{2}{\bf 1}_{4\times 4}&\\
\hline
&&(\alpha+\beta) (N+2) {\bf 1}_{p\times p}\\
\end{array}\right) \;.
\ee
Making the decomposition of the fields in the direct sum of representations in  the subgroup one gets Table~\ref{brsuv}.
\begin{table}[h!t]
{  \centering 
  \small{\begin{tabular}{|c|c |c|c|c|c|  }
\hline
\su  &   $SU(N)_{{\rm cf}_{\eta}}   $    &  $SU(4)_{\eta}$     &  $ SU(p)_{\eta\xi}$ &   $  U(1)_{\psi \eta}^{\prime}$   &   $ U(1)_{\psi \xi}^{\prime}$ \\
 \hline
  \sbu $\psi$   &   $ { \yng(2)} $  &    $  \frac{N(N+1)}{2} \cdot   (\cdot) $    & $ \frac{N(N+1)}{2} \cdot   (\cdot)   $   &  $N +4 +p$ &  $p $  \\
   $ \eta_1$      &   $  {\bar  {\yng(2)}} \oplus {\bar  {\yng(1,1)}}  $     & $N^2  \cdot  (\cdot )  $     &   $ N^2  \cdot  (\cdot ) $    &$-(N +4 +p)$ & $-p $\\
    $ \eta_2$      &   $ 4  \cdot   {\bar  {\yng(1)}}   $     & $N  \cdot  {\yng(1)}  $     &   $ 4 N  \cdot  (\cdot ) $   & $-\frac{N+ p + 4}{2}$  & $-\frac{p}{2}$\\
  $ \eta_3$      &   $ p  \cdot   {\bar  {\yng(1)}}$     & $ N p  \cdot  (\cdot )  $     &   $N  \cdot  \bar{\yng(1)}  $    &$0$ &$N+2$ \\
    $ \xi$      &   $ p  \cdot   { {\yng(1)}}   $     & $N p  \cdot  (\cdot )  $     &   $ N  \cdot   { {\yng(1)}}   $   & $0$  & $-(N+2)$\\
\hline 
\end{tabular}}  
  \caption{\footnotesize   UV fieds in the BY model, decomposed as a direct sum of the representations of the unbroken group of  Eq.~(\ref{symbresBis}). }
\label{brsuv}
}
\end{table}
\begin{table}[h!t]
{  \centering 
  \small{\begin{tabular}{|c|c |c|c|c|c|  }
\hline
 \su   &  $SU(N)_{{\rm cf}_{\eta}}   $    &  $SU(4)_{\eta}$     &  $ SU(p)_{\eta\xi}$ &   $ U(1)_{\psi \eta}^{\prime}  $   &   $ U(1)_{\psi \xi}^{\prime}$ \\
 \hline
      \sbbuu $ { B}_{1a} $      &  $ {\bar  {\yng(1,1)}}   $         &  $  \frac{N(N-1)}{2} \cdot  (\cdot) $        &    $  \frac{N(N-1)}{2} \cdot  (\cdot) $     & $-(N +4 +p)$  &  $ -p $\\
   $ { B}_{1b}  $      &  $   4 \cdot {\bar  {\yng(1)}}   $         &  $N  \cdot  {\yng(1)}  $        &    $4 N  \cdot  (\cdot ) $   & $-\frac{N+ p + 4}{2}$  & $-\frac{p}{2}$\\
      $ { B}_{1c}  $      &  $   6 \cdot (\cdot )    $         &  $ {\yng(1,1)}  $        &    $  6   \cdot  (\cdot ) $   & $0$  & $0$\\
\hline
\end{tabular}}  
  \caption{\footnotesize    IR massless fermions in the BY model  in the Higgs phase. }
\label{brsir}
}
\end{table}
The composite massless baryons are subset of  those in  (\ref{baryons10}):
\bea
 && {B}_{1a}^{[AB]} =  \psi^{ij}   \eta_{i}^{A}  \eta_{j}^{B} \;,  \qquad 
 {B}_{1b}^{[AC]} =  \psi^{ij}   \eta_{i}^{A}  \eta_{j}^{C}\;,  \qquad 
 {B}_{1c}^{[CD]} =  \psi^{ij}   \eta_{i}^{C}  \eta_{j}^{D}\;,  \nn \\
&&   A,B = 1, \dots, N \;,  \quad C,D=N+1, \dots, N+4 \;.       \label{specialbaryons}  
\eea
It is quite straightforward (see the remark in Introduction)  to verify that the UV-IR  anomaly matching continues to work, with  the UV fermions in Table~\ref{brsuv}
and the IR fermions in Table~\ref{brsir}.  

As in the $\psi\eta$ model,   the  baryons indicated as  $B_{1c}$ in Table~\ref{brsir}  were not considered in the earlier work on the BY models \cite{ADS,BKL4},   but assumed here  to be present and  massless.  These extra baryons do not contribute  to the triangle anomalies  with respect to unbroken symmetry group, see Table~\ref{brsir}, therefore  do not affect the  anomaly matching argument.  
However, as in the $\psi\eta$ model,  the condensate $  \brc     { B}_{1c}  { B}_{1c}   \ckt \sim  \brc  \psi \eta \eta\,  \psi \eta \eta \ckt $  is needed
 in order to reproduce the strong anomaly in the IR correctly.  This however does  not alter  neither the symmetry breaking pattern (\ref{symbresBis}), nor the conventional  anomaly matching. See Sec.~\ref{BY} for more discussions.

\section{Confining phase with unbroken global symmetries  of the  GG  models     \label{conf4}}

The candidate massless composite fermions  for the Georgi-Glashow models are:
  \be    
 {({ B}_{1})}^{\{AB\}}=    \chi^{ij}  \eta_i^{A}  \eta_j^{B} \;,
\qquad {({ B}_{2})}^{a}_{A}=    \bar{\chi}_{ij}  \bar{\eta}^{i}_{A}  \xi^{j,a} \;,
\qquad {({ B}_{3})}_{[ab]}=    \chi^{ij}  \bxi_{i,a}  \bxi_{j,b} \;,
\label{baryons20}
\ee
the first  symmetric in $A \leftrightarrow B$ and the third anti-symmetric in $a \leftrightarrow b$.
Writing the spin indices explicitly  they are: 
 \bea   
 & {({ B}_{1})}^{\{AB\}, \alpha}=  \frac{1}{2}   \epsilon_{\beta\gamma}  \, \chi^{ij, \beta}  \eta_i^{A, \gamma}  \eta_j^{B, \alpha}  +
  \frac{1}{2}   \epsilon_{\beta\gamma}  \, \chi^{ij, \beta}  \eta_i^{A, \alpha}  \eta_j^{B, \gamma}    \;,    
\nn \\    &{({ B}_{2})}^{a, \alpha}_{A}=     \epsilon_{\dot{\beta} \dot{\gamma}} \   \bar{\chi}_{ij}^{\dot \beta}     \bar{\eta}^{i, \dot{\gamma}}_{A}  \xi^{j,a, \alpha} \;,  \qquad    {({ B}_{3})}_{[ab]}=   \epsilon_{\dot{\beta} \dot{\gamma}} \       \chi^{ij}  \bxi_{i,a}^{\dot \beta}  \bxi_{j,b}^{\dot{\gamma}}  
\;. &
\label{baryons1}
\eea
All anomaly triangles  are saturated  by these candidate massless composite fermions,  see Table~\ref{suvsirBis}.
\begin{table}[h!t]
  \centering 
  \small{\begin{tabular}{|c|c|c |c|c| c|c| }
\hline
\su      &  $SU(N)_{\rm c}  $    &  $ SU(N-4+p)$    &  $ SU(p)$     &   $ {U}(1)_{\chi\eta}   $  &   $ {U}(1)_{\chi\xi}   $  \\
  \hline     
 \sbbu      $ { B}_{1}$   &  $  \frac{(N-4+p)(N-3+p)}{2} \cdot ( \cdot )    $         &  $ {\yng(2)}$        &    $  \frac{(N-4+p)(N-3+p)}{2} \cdot ( \cdot )    $   &  $-N+p $ & $p$ \\
  \hline     
 \sbbu    $ { B}_{2}$  &   $ (N-4+p) p \cdot ( \cdot )  $   &        $ p \cdot  \bar{\yng(1)}$   &     $  (N-4+p) \cdot  {\yng(1)}$     & $-(p-2)$ & $-(N+p-2)$\\
  \hline     
   \sbbuu  $ { B}_{3}$   &    $  \frac{p (p-1)}{2}  \cdot ( \cdot )  $  &      $ \frac{p (p-1)}{2}  \cdot ( \cdot )  $ &    $ \bar{\yng(1,1)}$   & $N-4+p$ & $2N -4 +p$\\
\hline
\end{tabular}}
  \caption{\footnotesize  IR massless fermions in the chirally symmetric phase of the  GG  model. 
}\label{air}
\end{table}
\begin{table}[h!t]
\centering
 \tiny{\begin{tabular}{|c|c|c|}
\hline
\su      &  UV    &  IR  \\
\hline 
\su $ SU(N-4+p)^3$     &   $ N $     &  $ N+p -p $ \\
\su $ SU(p)^3$     &   $ N $     &  $ N-4+p -(p-4) $ \\
\su $ SU(N-4+p)^2 - {U}(1)_{\chi\eta}$     &   $ - N(N-2)  $     &  $ -(N-2+p)(N-p) -p(p-2) $ \\
\su $ SU(N-4+p)^2- {U}(1)_{\chi\xi}$     &    $ 0 $     &  $  (N-2+p)p -p(N+p-2)  $ \\
\su $ SU(p)^2- {U}(1)_{\chi\eta}$     &   $ 0 $     &  $ -(N-4+p)(p-2) +(p-2)(N-4+p)  $ \\
\su $ SU(p)^2- {U}(1)_{\chi\xi}$     &   $ - N(N-2)   $     &  $   -(N-4+p)(N+p-2) +(p-2)(2N-4+p +0)$ \\
\su $ {U}(1)_{\chi\eta}^3$     &   $ \frac{N(N-1)}{2}(N-4+p)^3 - N(N-4+p) (N-2)^3  $     &  $ - \frac{(N-4+p)(N-3+p)}{2} (N-p)^3 -(N-4+p) p (p-2)^3 + $ \\\
\su      &        &  $ +  \frac{p (p-1)}{2}(N-4+p)^3$ \\
\su $ {U}(1)_{\chi\xi}^3$     &   $ \frac{N(N-1)}{2}p^3 - N p (N-2)^3  $     &  $    \frac{(N-4+p)(N-3+p)}{2} p^3 -(N-4+p) p (N+p-2)^3  +$ \\
\su    &      &  $    \  +  \frac{p (p-1)}{2}(2N-4+p)^3$ \\
\su $ {\rm Grav}^2-{U}(1)_{\chi\eta} $     &  $ \frac{N(N-1)}{2}(N-4+p)  - N(N-4+p) (N-2)   $     &  $ - \frac{(N-4+p)(N-3+p)}{2} (N-p)  -(N-4+p) p (p-2) + $ \\
\su      &       &  $ +  \frac{p (p-1)}{2}(N-4+p) $ \\
\su $ {\rm Grav}^2-{U}(1)_{\chi\xi}$       &   $ \frac{N(N-1)}{2}p - N p (N-2) $     &  $    \frac{(N-4+p)(N-3+p)}{2} p -(N-4+p) p (N+p-2)  +$ \\
\su    &      &  $    \  +  \frac{p (p-1)}{2}(2N-4+p)$ \\
\su $ SU(N-4+p)^2-({\mathbbm Z}_{N-2})_{\chi}  $     &  $0$    & $N-2+p-p = 0 \ {\rm mod} \ N-2$    \\
\su $ SU(p)^2- ({\mathbbm Z}_{N-2})_{\chi} $     &   $0$    & $-(N-4+p) + p-2  = 0 \ {\rm mod} \ N-2 $     \\
\su $ {\rm Grav}^2-({\mathbbm Z}_{N-2})_{\chi}  $     &  $1$      & $1-1+1$  \\
\hline   
\end{tabular}}
  \caption{\footnotesize  Anomaly matching checks for the IR chiral symmetric phase of the GG  model.
}\label{suvsirBis}
\end{table}

\section{Dynamical Higgs phase in the generalized GG  models     \label{Higgs4} }

In  the   generalized Georgi-Glashow  models there is  a competition between two possible bifermion symmetry breaking channels $\chi \eta$ and $  \xi \eta $. 
This time,   the MAC criterion would slightly   favor the  $  \xi \eta $  condensates against  $\chi \eta$. Indeed,      
the strength of the one-gluon exchange potential for the two channels
 \bea   &     \chi \left({\raisebox{-9pt}{\yng(1,1)}}\right) \,  \,  \eta \Big(\bar{\raisebox{-2pt}{\yng(1)}}\Big)      \qquad  \   & {\rm forming}    \quad      \raisebox{-2pt}{\yng(1)}\;,\nonumber    \\
&     \xi \Big({\raisebox{-2pt}{\yng(1)}}\Big) \, \eta \Big(\bar{\raisebox{-2pt}{\yng(1)}}\Big)    \qquad \ & {\rm forming}  \quad  (\cdot)  \;,
 \eea
are, respectively,
 \bea       \frac{N^2-1}{2N}  -      \frac{ (N-2)(N+1)}{N}  -   \frac{N^2-1}{2N}  &=&  -      \frac{ (N-2)(N+1)}{N} \;,\nonumber \\
     0 -   2       \frac{N^2-1}{2N}   & =&  -       \frac{N^2-1}{N}  \;.
 \eea
Again, these perturbative estimates are not excessively significant, and  we assume that
 both  condensates occur as:
\bea
&& \brc  \chi^{ij}   \eta_i^A \ckt  =\,   c_{\chi\eta} \,  \Lambda^3   \delta^{j A}\ne 0\;,   \qquad   j=1,\dots,  N-4\;,   \quad  A=1, \dots , N -4 \;,  \nonumber \\
&&  \brc  \xi^{i,a}   \eta_i^B \ckt =\,   c_{\eta\xi} \,  \Lambda^3   \delta^{aB}\ne 0\;,   \qquad  a=1,\dots,  p \;,   \quad B=N-4+1,\dots,N-4+  p \;. \nonumber \\ 
  \eea
The pattern of symmetry breaking is
\bea
&& SU(N)_{\rmc}  \times   SU(N-4+p)_{\eta}  \times  SU(p)_{\xi}  \times  U(1)_{\chi\eta}\times  U(1)_{\chi\xi} \nn \\
&&  \xrightarrow{\brc  \xi   \eta \ckt , \brc  \chi \eta \ckt}      SU(4)_{{\rm  c}}  \times  SU(N-4)_{{\rm cf}_{\eta}}  \times   SU(p)_{\eta\xi}  \times  U(1)_{\chi \eta}^{\prime} \times  U(1)_{\chi \xi}^{\prime} \;.
\label{symbrea}
\eea

The color gauge symmetry is partially (dynamically) broken, leaving  color-flavor diagonal  global $SU(N-4)_{{\rm cf}_{\eta}} $ symmetry and an $SU(4)_{{\rm  c}}$ gauge symmetry.
$U(1)_{\chi \eta}^{\prime}$ and $U(1)_{\chi \xi}^{\prime}$ are a combinations respectively  of $U(1)_{\chi \eta}$ and $U(1)_{\chi \xi}$ with  the elements of $SU(N)_{\rmc} $ and $ SU(N-4+p)_{\eta}$ generated by:
\bea  
&& \qquad  t_{SU(N)_{\rm c}}=  \left(\begin{array}{c|c}
2 \frac{  \alpha  (N-4+p ) +\beta p }{N-4}  {\bf 1}_{(N-4)\times (N-4)}&\\
\hline
& - \frac{  \alpha  (N-4+p ) +\beta p }{2}  {\bf 1}_{4\times 4}\\
\end{array}\right) \;, \nn \\
&& t_{SU(N-4+p)_{\eta}}=  \left(\begin{array}{c|c}
-\frac{p (\alpha + \beta ) (N-2 )}{N-4} {\bf 1}_{(N-4)\times (N-4)}&\\
\hline
&  (\alpha + \beta ) (N-2 )  {\bf 1}_{p\times p}\\
\end{array}\right) \;.  
\eea
Making the decomposition of the fields in the direct sum of representations in  the subgroup one arrives at  Table~\ref{brauv}.
\begin{table}[h!t]
{  \centering 
  \small{\begin{tabular}{|c|c |c|c|c|c|  }
\hline
\su       &    $SU(N-4)_{{\rm cf}_{\eta}}   $  &  $SU(4)_{{\rm c}}$ &  $ SU(p)_{\eta\xi}$ &   $  U(1)_{\chi \eta}^{\prime}$   &   $ U(1)_{\chi \xi}^{\prime}$ \\
\hline
\sbbuu $\chi_1$     &       $ { \yng(1,1)} $ & $  \frac{(N-4)(N-5)}{2} \cdot   (\cdot) $  & $ \frac{(N-4)(N-5)}{2} \cdot   (\cdot)   $   &  $\frac{(N-4+p) N}{(N-4)}    $  &  $p \frac{N}{N-4}$  \\
$\chi_2$                &   $ 4 \cdot { \yng(1)} $  &    $  (N-4)  \cdot    { \yng(1)} $    & $ 4(N-4) \cdot   (\cdot)   $   &$\frac{(N-4+p) N}{2(N-4)}  $  &  $\frac{p N}{2(N-4)}$  \\
$\chi_3$                &    $  6  \cdot   (\cdot) $   &   $ { \yng(1,1)} $   & $ 6  \cdot   (\cdot)   $   &  $0$  & $0$  \\
$ \eta_1$                 &  $  {\bar  {\yng(2)}} \oplus {\bar  {\yng(1,1)}}  $     &   $(N-4)^2  \cdot   (\cdot )    $   &   $ (N-4)^2  \cdot  (\cdot ) $   & $ -\frac{(N-4+p) N}{(N-4)} $  & $-\frac{p N}{N-4} $\\   
$ \eta_2$             & $ p  \cdot \bar{\yng(1)} $      &   $p(N-4)  \cdot  (\cdot )  $      &   $ (N-4)  \cdot \bar{\yng(1)} $    &$-2 -2 \frac{p}{N-4}$ & $N-2 -\frac{2 p}{N-4}  $\\
$ \eta_3$              &   $  4   \cdot   {\bar  {\yng(1)}}$     & $ (N-4)  \cdot   {\bar  {\yng(1)}} $     &   $ 4  (N-4)     \cdot   (\cdot )  $    &$-\frac{(N-4+p) N}{2(N-4)} $ &$ -\frac{p N}{2(N-4)} $ \\
$ \eta_4$              &   $ 4 p \cdot  (\cdot ) $     & $ p \cdot   {\bar  {\yng(1)}}   $     &   $4  \cdot \bar{\yng(1)}   $    &$\frac{N-4+p }{2}$ &$N-2  + \frac{p }{2}$ \\
$ \xi_1$                &   $  p  \cdot   { {\yng(1)}}    $     & $p(N-4)   \cdot  (\cdot )  $     &   $ (N-4)   \cdot   { {\yng(1)}}   $   & $2 + 2\frac{p}{N-4}$  & $-(N-2)+\frac{2p}{N-4}$\\
$ \xi_2$                &   $4 p   \cdot  (\cdot )   $     & $  p  \cdot   { {\yng(1)}}    $     &   $ 4  \cdot   { {\yng(1)}}   $   & $-\frac{N-4+p}{2}$  & $-(N-2)-\frac{p}{2}$\\
\hline 
\end{tabular}}  
  \caption{\footnotesize   UV fieds in the GG model, decomposed as a direct sum of the representations of the unbroken group of  Eq.~(\ref{symbrea}). }
\label{brauv}
}
\end{table}

\begin{table}[h!t]
{  \centering 
  \small{\begin{tabular}{|c|c |c|c|c|  }
\hline
\su     &    $SU(N-4)_{{\rm cf}_{\eta}}   $    &  $ SU(p)_{\eta\xi}$ &   $  U(1)_{\chi \eta}^{\prime}$   &   $ U(1)_{\chi \xi}^{\prime}$ \\
 \hline
\sbbu $ { B}   $          &  $  {\bar  {\yng(2)}}  $     &   $\frac{(N-4)(N-3)}{2}\  \cdot  (\cdot ) $   &  $ -\frac{(N-4+p) N}{(N-4)} $  & $-\frac{p N}{N-4} $\\
  \hline 
\end{tabular}}  
  \caption{\footnotesize   IR fied in the GG model in the dynamical Higgs phase. }
\label{brair}
}
\end{table}
The composite massless baryons are subset of those in (\ref{baryons20}): 
\bea
 && { B}^{\{AB\}} =  \chi^{ij}   \eta_{i}^{A}  \eta_{j}^{B} \;,  \qquad  A,B = 1, \dots, N-4 \;.   \label{baryonbra}
\eea
In the IR these fermions  saturate all the anomalies of the unbroken chiral symmetry. This can be seen by  an inspection of  Table~\ref{brair} and   Table~\ref{brauv},    with the help of the following observation.   

In fact,   there is a novel feature in the  GG models, which is not shared by the  BY models.   As  seen in  Table~\ref{brair},
there is an  unbroken strong gauge symmetry $SU(4)_{{\rm c}}$, with a  set of  fermions, 
\be  \chi_3\;,\quad \chi_2\;,\quad  \eta_3\;,\quad \eta_4\;,\quad \xi_2\;,\label{these}
\ee
charged with respect to it.     However,   the pairs  $\{ \chi_2\;,\,  \eta_3 \}$  and   $\{ \eta_4\;,\,  \xi_2 \}$ can form massive Dirac fermions and decouple.  These are vectorlike  with respect to the  surviving infrared symmetry, (\ref{symbrea}), hence are irrelevant to the anomalies.\footnote{Actually,   
  with matter fermions (\ref{these})  $SU(4)_{{\rm c}}$  is asymptotically free only for    $50- 2N - 2 p  > 0$.   If     $50- 2N - 2 p  <  0$,    $SU(4)_{{\rm c}}$  will remain weakly coupled in the infrared, 
but the fact that the fermions (\ref{these}) do not contribute to the anomalies with respect to  the remaining flavor symmetries  (\ref{remainingfl})  stays valid.}
   On the other hand,   the fermion $\chi_3$  can condense 
\be    \brc  \chi_3 \chi_3 \ckt  \
\ee
forming massive composite mesons, $\sim \chi_3 \chi_3$, which also decouples.  It is again neutral with respect to all of 
\be SU(N-4)_{{\rm cf}_{\eta}}  \times   SU(p)_{\eta\xi}  \times  U(1)_{\chi \eta}^{\prime} \times  U(1)_{\chi \xi}^{\prime} \;.\label{remainingfl}
\ee
To summarize,    $SU(4)_{\rm c}$  is invisible  (confines) in the IR, and  only the unpaired  part of the $\eta_1$  fermion   $\big({\bar  {\yng(2)}}\big)$  remains  massless, and its contribution to the anomalies is reproduced exactly by the composite fermions, (\ref{baryonbra}).

{\noindent  \bf Comment:}
The massive mesons  $\{ \chi_2\,  \eta_3 \}$,    $\{ \eta_4 \,  \xi_2 \}$,  $ \{\chi_3 \, \chi_3\}$  are not charged with respect to the flavor symmetries surviving in the infrared.  
It is tempting to regard them as a toy-model  ``dark matter",  as contrasted to  the fermions  $ { B}^{AB} $ which constitute the ``ordinary, visible" sector.

\end{document}